%% file: main.tex
\documentclass[acmsmall]{acmart}
\newbool{showComments}
\boolfalse{showComments}    % UNCOMMENT THIS BEFORE SUBMISSION
\ifbool{showComments}{
\newcommand{\chulhong}[1]{\textcolor{blue}{\textbf{cm:} #1}}
\newcommand{\cm}[1]{\textcolor{blue}{\textbf{cm:} #1}}
\newcommand{\seungwoo}[1]{\textcolor{blue}{\textbf{sw:} #1}}
\newcommand{\euihyeok}[1]{\textcolor{brown}{\textbf{Euihyeok:} #1}}
\newcommand{\dongwoo}[1]{\textcolor{blue}{\textbf{dongwoo:} #1}}
\newcommand{\jin}[1]{\textcolor{blue}{\textbf{Jin:} #1}}

\newcommand{\highlight}[1]{\textcolor{red}{#1}}
\newcommand{\revise}[1]{\textcolor{red}{#1}}

\newcommand{\needtofill}[1]{\textcolor{green}{\textbf{#1}}}
\newcommand{\deleteifnospace}[1]{\textcolor{brown}{#1}}
}{
\newcommand{\chulhong}[1]{}
\newcommand{\cm}[1]{}
\newcommand{\seungwoo}[1]{}
\newcommand{\euihyeok}[1]{}
\newcommand{\dongwoo}[1]{}
\newcommand{\jin}[1]{}

\newcommand{\highlight}[1]{#1}
\newcommand{\revise}[1]{#1}
\newcommand{\needtofill}[1]{}

\newcommand{\deleteifnospace}[1]{}
}

\usepackage{kotex}
\usepackage[ampersand]{easylist}
\usepackage{booktabs} % For formal tables
\usepackage{import}
\usepackage[ruled]{algorithm2e} % For algorithms
\usepackage[utf8]{inputenc}

\usepackage[symbol]{footmisc} % to workaround footnote in title
\usepackage{booktabs,tabularx}
\usepackage{subcaption}
\usepackage{adjustbox}
\usepackage{newunicodechar}
\usepackage{multirow}
\usepackage{enumitem}
\usepackage{graphicx,subcaption}
\usepackage{hhline}
\usepackage{comment}
\usepackage{makecell}
\usepackage{caption}
\AtBeginDocument{%
  \providecommand\BibTeX{{%
    \normalfont B\kern-0.5em{\scshape i\kern-0.25em b}\kern-0.8em\TeX}}}

\setcopyright{acmcopyright}
\copyrightyear{2018}
\acmYear{2018}
\acmDOI{10.1145/1122445.1122456}

\acmConference[Woodstock '18]{Woodstock '18: ACM Symposium on Neural
  Gaze Detection}{June 03--05, 2018}{Woodstock, NY}
\acmBooktitle{Woodstock '18: ACM Symposium on Neural Gaze Detection,
  June 03--05, 2018, Woodstock, NY}
\acmPrice{15.00}
\acmISBN{978-1-4503-XXXX-X/18/06}

\begin{document}

%%
%% The "title" command has an optional parameter,
%% allowing the author to define a "short title" to be used in page headers.
\title{Automatic Detection of Reactions to Music via Earable Sensing}

%%
%% The "author" command and its associated commands are used to define
%% the authors and their affiliations.
%% Of note is the shared affiliation of the first two authors, and the
%% "authornote" and "authornotemark" commands
%% used to denote shared contribution to the research.

% \author{Anonymous authors}

\author{Euihyeok Lee}
\email{euihyeok.lee@misl.koreatech.ac.kr}
\affiliation{%
  \institution{KOREATECH}
  \country{Republic of Korea}
}

\author{Chulhong Min}
\email{chulhong.min@nokia-bell-labs.com}
\affiliation{%
  \institution{Nokia Bell Labs}
  \country{UK}
}

\author{Jaeseung Lee}
\email{jaeseung.lee@misl.koreatech.ac.kr}
\affiliation{%
  \institution{KOREATECH}
  \country{Republic of Korea}
}

\author{Jin Yu}
\email{jin.yu@misl.koreatech.ac.kr}
\affiliation{%
  \institution{KOREATECH}
  \country{Republic of Korea}
}

\author{Seungwoo Kang}
%\authornote{Corresponding author}
% \authornotemark[1]
\email{swkang@koreatech.ac.kr}
\affiliation{%
  \institution{KOREATECH}
  \country{Republic of Korea}
}

% \settopmatter{printacmref=false}

%%
%% By default, the full list of authors will be used in the page
%% headers. Often, this list is too long, and will overlap
%% other information printed in the page headers. This command allows
%% the author to define a more concise list
%% of authors' names for this purpose.
%\renewcommand{\shortauthors}{Trovato and Tobin, et al.}
\renewcommand{\shortauthors}{Lee, et al.}

%%
%% The abstract is a short summary of the work to be presented in the
%% article.
\input{sections/00.abstract.tex}

%%
%% The code below is generated by the tool at http://dl.acm.org/ccs.cfm.
%% Please copy and paste the code instead of the example below.
%%


% \ccsdesc[300]{Computer systems organization~Embedded systems}
% \ccsdesc[500]{Human-centered computing~Ubiquitous and mobile computing systems and tools}

%%
%% Keywords. The author(s) should pick words that accurately describe
%% the work being presented. Separate the keywords with commas.

\settopmatter{printacmref=false}
\renewcommand\footnotetextcopyrightpermission[1]{} % removes footnote with conference information in first column
\pagestyle{plain} % removes running headers


% \renewcommand\footnotetextcopyrightpermission[1]{} % removes footnote with conference information in first column
% \pagestyle{plain} % removes running headers

%%
%% This command processes the author and affiliation and title
%% information and builds the first part of the formatted document.
\maketitle

\input{sections/01.introduction.tex}
\input{sections/02.relatedwork.tex}
\input{sections/03-new.overview}

\input{sections/04.system}
\input{sections/041.vocal}

\input{sections/042.motion}

\input{sections/06.datacollection}

\input{sections/07.evaluation}

\input{sections/08.discussion}
\input{sections/09.conclusion}

%%
%% The acknowledgments section is defined using the "acks" environment
%% (and NOT an unnumbered section). This ensures the proper
%% identification of the section in the article metadata, and the
%% consistent spelling of the heading.
%\begin{acks}
%To Robert, for the bagels and explaining CMYK and color spaces.
%\end{acks}

%%
%% The next two lines define the bibliography style to be used, and
%% the bibliography file.

\bibliographystyle{ACM-Reference-Format}
\bibliography{reference}

%%
%% If your work has an appendix, this is the place to put it.
%\appendix
%\section{Research Methods}

\end{document}

%% file: sections/00.abstract.tex
\begin{abstract}

We present GrooveMeter, a novel system that automatically detects vocal and motion reactions to music via earable sensing and supports music engagement-aware applications. To this end, we use smart earbuds as sensing devices, which are already widely used for music listening, and devise reaction detection techniques by leveraging an inertial measurement unit (IMU) and a microphone on earbuds. To explore reactions in daily music-listening situations, we collect the first kind of dataset, MusicReactionSet, containing 926-minute-long IMU and audio data with 30 participants. With the dataset, we discover a set of unique challenges in detecting music listening reactions accurately and robustly using audio and motion sensing. We devise sophisticated processing pipelines to make reaction detection accurate and efficient. We present a comprehensive evaluation to examine the performance of reaction detection and system cost. It shows that GrooveMeter achieves the macro $F_1$ scores of 0.89 for vocal reaction and 0.81 for motion reaction with leave-one-subject-out cross-validation. More importantly, GrooveMeter shows higher accuracy and robustness compared to alternative methods. We also show that our filtering approach reduces 50\% or more of the energy overhead. Finally, we demonstrate the potential use cases through a case study.

\end{abstract}

%% file: sections/01.introduction.tex
\section{Introduction}

\begin{figure}[t]
    \centering
    \mbox{
        \hspace{-0.2in}
        \subfloat[Music player\label{fig:playing_screenshot}]{\includegraphics[width=0.23\textwidth]{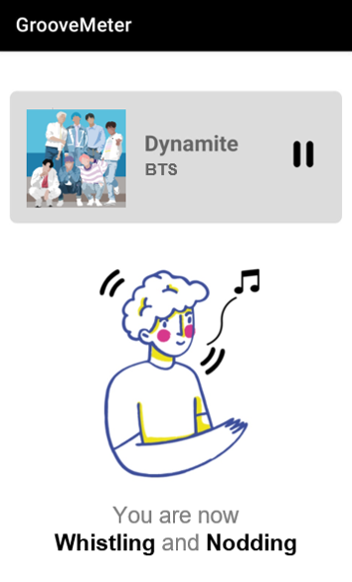}}
        \hspace{0.05in}
        \subfloat[Automatic rating\label{fig:reaction_detection_screenshot}]{\includegraphics[width=0.23\textwidth]{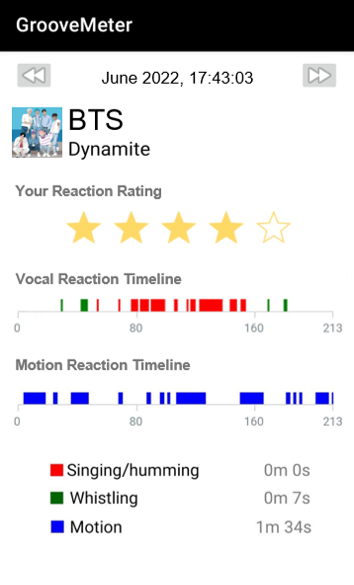}}
        
        \hspace{0.05in}
        \subfloat[Recommendation\label{fig:recommendation_screenshot}]{\includegraphics[width=0.23\textwidth]{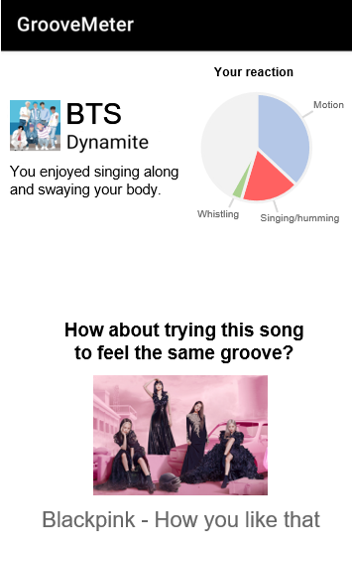}}
        }
  
    \caption{Music engagement-aware applications built on GrooveMeter ~\label{fig:gm_app}}
 
\end{figure}

Listening to music is an integral part of our life. According to a study~\cite{ml21}, in 2021, music consumers have listened to music for more than 2.63 hours daily, which is equivalent to listening to 53 songs every day. %\chulhong{Updated numbers with the 2021 report.} 
While people listen to songs or music, they often nod their head, tap their foot, hum or sing along to the songs at the same time. These are the natural responses of people listening to music~\cite{pawley2012science, burger2013influences, burger2014hunting, levitin2018psychology}, which are considered as a characteristic showing their engagement with music~\cite{godoy2010musical, agrawal2020towards}. These reactions are compelling to enable interesting \emph{music engagement-aware applications}. For example, music player apps (Figure~\ref{fig:playing_screenshot}) can leverage listeners' on-the-fly reactions to provide an engagement-aware automatic music rating (Figure~\ref{fig:reaction_detection_screenshot}) and reaction-based music recommendation (Figure~\ref{fig:recommendation_screenshot}), by observing which part of a song listeners often reacted to while listening.

Observing responses to music listening has widely been investigated in music psychology studies. They provided insightful findings to understand the characteristics of responses to music listening, but it is almost impossible to adopt these methods for real-life applications because they mostly rely on self-report or bulky experimental equipment in a controlled environment. For example, the previous studies measured brain activity using positron emission tomography (PET) or functional magnetic resonance imaging (fMRI)~\cite{blood2001intensely, koelsch2006investigating}, observed music-induced movement using motion capture systems~\cite{burger2013influences, luck2010effects, gonzalez2018correspondences}, and measured physiological responses such as an electrocardiogram (ECG) and galvanic skin response (GSR)~\cite{lynar2017joy}.

In this paper, we propose \emph{GrooveMeter}, a novel mobile system that tracks reactions to music listening and supports music engagement-aware applications. While there are various types of music reactions, as an initial attempt, we focus on external and readily observable bodily reactions, i.e., \emph{physical responses}, which people usually experience while listening to music~\cite{hallam2016oxford}. Specifically, we target singing along, humming, whistling, and head motion because they are common reactions observed from our in-the-wild dataset presented in \S\ref{sec:collection}. To this end, we use smart earbuds as sensing devices, which are already widely used for music listening, and devise reaction detection techniques by leveraging an inertial measurement unit (IMU) and a microphone on earbuds.

While significant research efforts have been made to recognize human activities and gestures over decades, detecting music listening reactions with sensor-equipped wearables has not been studied yet. Through extensive observation and analysis, we discover a set of unique challenges in detecting music listening reactions accurately and robustly using audio and motion sensing. First, there are often reaction-irrelevant events that show similar signal characteristics, which can cause false positive errors (e.g., mumbling to talk to him/herself, drinking coffee, or looking at a monitor and a keyboard alternately).
Second, since listening to music is often done as a secondary activity, audio and motion signals can be affected by background noise (e.g., a sound of nearby people talking or background music in a cafe) and other motion artifacts, respectively. Models trained with data from a lab environment suffer from serious performance degradation in diverse daily music-listening situations. 
Third, running sound and motion classification models on mobile devices incurs considerable processing and energy cost for continuous execution.

To address the challenges, we devise sophisticated processing pipelines for vocal and motion reaction detection with three main features. First, we investigate the signal characteristics of data segments that can be \textit{certainly} labeled as non-reaction and devise a filtering approach to effectively filter out those segments in the beginning of the processing pipeline, not only for cost-saving but also improving the robustness. 
Second, we elaborate multi-step reaction detection pipelines, reflecting the unique patterns of music listening reactions.  
Third, we leverage the semantic similarity between sensor data from a listener and \emph{musical structure information} retrieved from a song. The intuition is that music listening reactions show a high correlation with the playing song. For example, a listener's humming would naturally follow the song's rhythm and pitch pattern. To this end, we correct ambiguously labeled audio segments based on the prosodic similarity at the last stage.

To build and evaluate GrooveMeter, we collect the first-kind-of dataset, called \emph{MusicReactionSet}, with 240 music listening sessions from 30 participants under four situations (resting in a lounge, working at an office, riding in a car, and relaxing at a cafe). It contains 926-minute-long IMU and audio data with manually-labeled accurate annotation. 

Our extensive evaluation using MusicReactionSet shows that GrooveMeter achieves the macro $F_1$ scores of 0.89 and 0.81 with leave-one-subject-out cross-validation for vocal and motion reaction detection, respectively. 
More importantly, GrooveMeter shows higher accuracy and robustness compared to alternative methods. Especially, in noisy situations, e.g., relaxing at a cafe, we observe a significant performance enhancement. Also, the filtering operation reduces 50\% or more of the energy overhead from our measurement.
In addition, to demonstrate the usefulness of GrooveMeter, we prototyped GrooveMeter on Android phones and smart earbuds and further developed a set of music engagement-aware applications on GrooveMeter (see Figure~\ref{fig:gm_app}). We demonstrate application case studies showing the feasibility of automatic music rating, music familiarity detection, and reaction-based music recommendation. 

We summarize the contribution of this paper as follows.
\begin{itemize}[noitemsep, topsep=4pt, leftmargin=*]
    \item We collect the first kind of dataset, MusicReactionSet, containing 926-minute-long IMU and audio data with 30 participants to explore vocal and motion reactions in daily music-listening situations. 
    \item We develop GrooveMeter, a novel system that detects reactions to music listening via earable sensing. To the best of our knowledge, this is the first to present an earable sensing solution specialized for automatic detection of vocal and motion reactions to music.
    \item We propose a novel technique to make reaction detection efficient and robust by filtering out reaction-irrelevant data segments and leveraging music information retrieved from a song, and present a comprehensive evaluation using MusicReactionSet.
\end{itemize}

%% file: sections/02.relatedwork.tex
\section{Related Work}

\textbf{Reaction sensing on content consumption:} 
There have been several attempts to monitor a consumer's reaction to multimedia content or performing arts, e.g., sensing the implicit responses of users watching movies~\cite{bao2013your}, measuring audience responses in live performances by monitoring spontaneous body movements~\cite{martella2015exploiting}, quantifying the experience of audiences in the play~\cite{wang2017play}, inferring humor appraisal of comic strips~\cite{barral2017no}, detecting frisson of audience during music performances based on physiological sensing~\cite{he2022frisson}, estimating the attention level of a learner in an online lecture from eye gaze and gaze gesture tracking~\cite{kar2020gestatten}. These works share a high-level goal with ours, aiming at detecting reactions of content consumers at runtime. However, due to different characteristics of content-dependent reactions, the required sensor modality, devices, and techniques should be different. We aim at detecting music listening reactions by reflecting unique signal characteristics of the reactions. We also discover an opportunity to make the detection robust in noisy conditions. Thus, we develop a novel technique to exploit the semantic similarity between sensor data and music information. Note that the previous works do not utilize the characteristic of contents for reaction sensing, but rely on the sensor data only.

\textbf{Human sensing using earables:}
Recent research efforts have tried to sense diverse human contexts using sensor-equipped earbuds. For example, IMU is used to detect physical activity~\cite{min2018exploring}, head motion~\cite{ferlini2019head}, facial expression~\cite{lee2019automatic,verma2021expressear}, respiration rate~\cite{roddiger2019towards}, jaw movement~\cite{khanna2021jawsense}, gait posture~\cite{jiang2022earwalk}, and mistakes in free weight exercises~\cite{radhakrishnan2020erica}. A microphone in earbuds is known to be able to capture the higher quality of a wearer's speech due to its close distance to the mouth~\cite{kawsar2018earables}. It can also detect interesting audio events made around a mouth, e.g., eating activity~\cite{min2018exploring}. Also, recent works present acoustic motion tracking ~\cite{cao2020earphonetrack}, gait sensing~\cite{ferlini2021eargate}, human activity recognition~\cite{ma2021oesense}, user authentication~\cite{wang2021eardynamic,xieteethpass}, tongue-jaw movement recognition~\cite{cao2021canalscan}, and facial expression tracking~\cite{li2022eario}. In-around-ear devices with physiological sensors have been developed for diverse purposes, e.g., facial expression detection~\cite{choi2022ppgface}, blood pressure monitoring~\cite{bui2019ebp}, microsleep detection~\cite{pham2020wake}. A recent work presents an earphone-based acoustic sensing system to detect ear diseases~\cite{jin2022earhealth}. Also,  MusicalHeart presents a system to continuously monitor the user's heart rate and activity level using a microphone and IMU in earphones~\cite{nirjon2012musicalheart}. In this work, we focus on music listening reactions and present novel sensing techniques to detect responses usually made while listening to music.

\textbf{Understanding music listening behavior and contexts:} 
Some previous studies tried to understand music listening behavior and contexts. One of the initial attempts with mobile technology was made in~\cite{north2004uses} in 2002. The authors used text messages to analyze the music people heard.
In~\cite{yang2015quantitative}, the authors developed a smartphone-based tool to collect music listening behavior, e.g., surrounding contexts, user activity, and mood. For contextual music recommendation, Volokhin et al. tried to understand the users' intent for listening to music and its relationship to common daily activities, but it relied on an online survey~\cite{volokhin2018understanding}. Our work differs from them in two aspects. While most of the existing studies focused on a user's behavior and contexts, i.e., what/when/why/where people listen to music, our work focuses on \emph{how people react} to music, which has been rarely studied in the mobile computing research. Next, we devise novel sensing and processing methods for reaction detection, thereby enabling large-scale data collection in real-life situations. We expect our work will complement existing research.

%% file: sections/03-new.overview.tex
\section{GrooveMeter Overview~\label{sec:reaction}}

\begin{figure}[t]
    \centering
    \includegraphics[width=0.70\columnwidth]{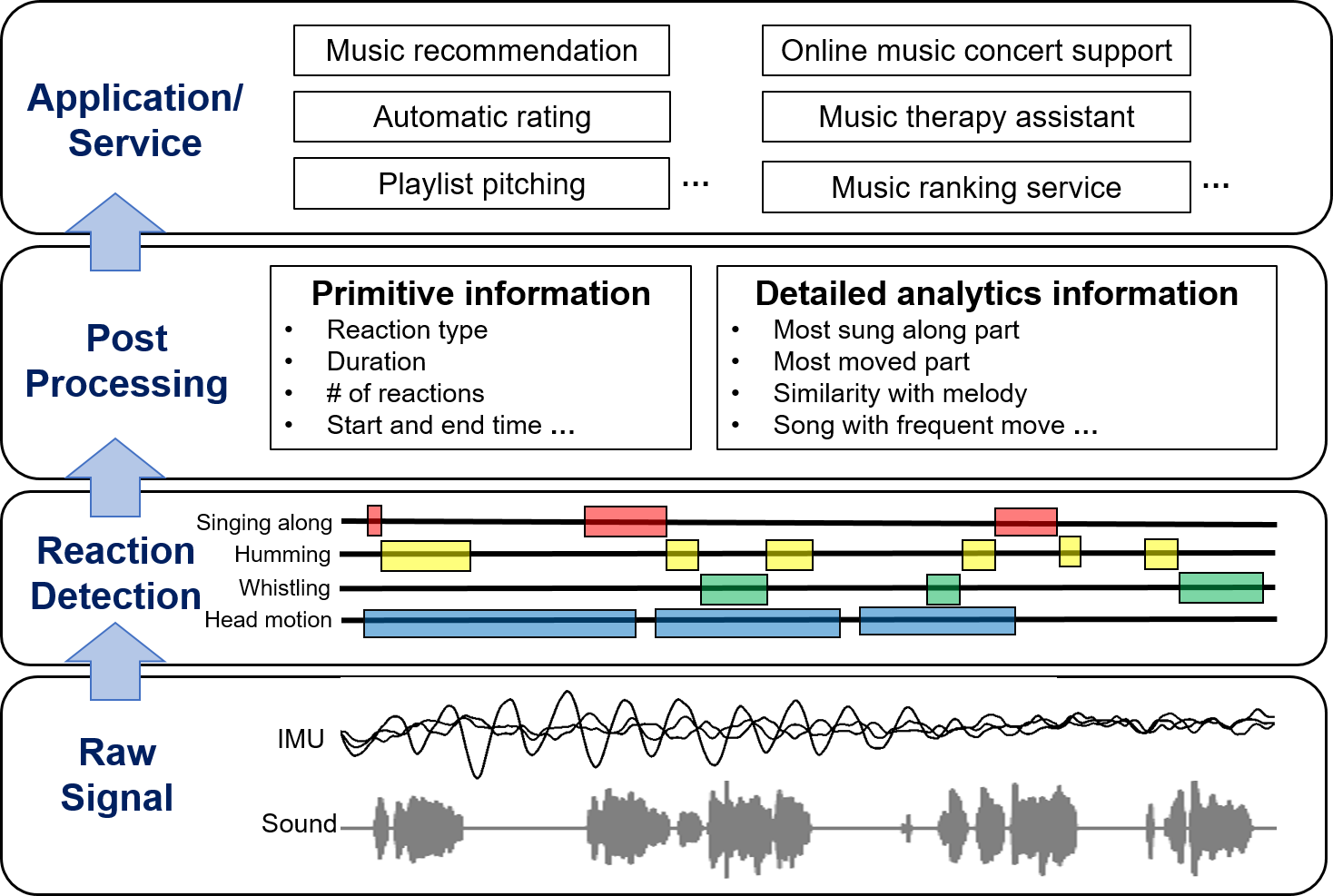}

    \caption{High-level process to support music engagement-aware applications. ~\label{fig:music_reaction}}
\vspace{-0.1in}
\end{figure}

\subsection{Example Applications on GrooveMeter}

Monitoring on-the-fly music listening reactions in unconstrained mobile environments opens a broad spectrum of applications. Figure~\ref{fig:music_reaction} shows the high-level process of how GrooveMeteter supports music engagement-aware applications. First, as a common basis for any applications, GrooveMeter focuses on detecting vocal and motion reactions in real time using IMU and audio signals, i.e., which type of reaction was made, and when and how long. By combining with these primitive information, GrooveMeter further provides music engagement-aware applications with high-fidelity information, e.g., which part of a song listeners most sung along or moved. In this work, we target singing along, humming and whistling as vocal reaction, and head motion as motion reaction. Note that these are commonly observed reactions from our in-the-wild dataset with 30 participants presented in \S\ref{sec:collection}.

In the rest of the subsection, we illustrate potential scenarios of music engagement-aware applications to show the usefulness of the reaction information provided by GrooveMeter.

\textbf{Automatic and fine-grained music rating:} One straightforward use case would be automatic and fine-grained music rating. Today's music rating is mostly relied on a user's manual input and simple statistics such as the number of plays to figure out a user's preference on songs. From the intuition that people would make different reaction patterns depending on how much they like a playing song, we envision that rating can be predicted with the reaction information. We develop the model for rating prediction using the reaction output from GrooveMeter and show its feasibility in \S\ref{subsec:casestudy}.

\textbf{Reaction-based music recommendation and playlist:} 
Generating customized and personalized playlists is important for music streaming service providers to attract more consumers and for consumers as well to easily discover music for their tastes from millions of songs. Spotify, one of the most popular music streaming services, provides a recommendation service based on a combination of collaborative filtering, natural language processing, and audio analysis models~\cite{spotify1, spotify2}. We envision that reaction information can extend the current strategy with additional user data, which show how listeners engage with the music they are listening to, e.g., songs that made listeners sing along or headbang, etc. 
Here we do not argue that reaction-based music recommendation is a more advanced approach or provides a better result than the recommendation service of Spotify. We prototyped the application for reaction-based music recommendation using GrooveMeter and conducted a case study in \S\ref{subsec:casestudy}.

\textbf{Enriching remote interaction between fans and musicians:} Recently, many musicians have performed online music concerts due to COVID-19~\cite{forbesLiveEvents}. We envision that GrooveMeter, integrated with online concert platforms, can enrich interactions between musicians and remote fans. It detects and collects the reactions of fans watching live performances online. Collected reactions could be used to generate a form of collective sonification or visualization provided to the musicians. For example, synthesized sound effects of cheering and singing along from the vocal reactions could be generated to convey the fans' enthusiasms. The reactions and the effect could be shared among the fans, which would help them feel more engaged and immersed in the concert.

\textbf{Assistance tool for music therapy:} We envision that, beyond enriching music players, GrooveMeter can serve as a base to develop an assistance tool in various areas where music is used as a medium, e.g., music therapy, music psychology, and education. For example, during a music therapy session in the clinic, a therapist and a client do music experiences such as listening to music, singing, playing instruments, and creating music~\cite{bonde2002comprehensive}. To explore the potential of GrooveMeter in music therapy, we consulted a music therapist about current practices of music therapy out of the clinic and their limitations. The therapist mentioned that clients are sometimes given assignments so that they can apply the activities performed in the clinic session to the daily life. However, it is difficult for therapists to check the client's compliance completely because they currently rely on the checklist or verbal report by the clients. The therapist suggested that GrooveMeter could be a tool that provide objective measures to therapist by monitoring and analyzing clients' music listening sessions at home. GrooveMeter can be further extended to employ other physiological sensors to analyze clients' mental status, e.g., emotion and stress.

\subsection{Design Considerations}

We propose GrooveMeter, a system to track a user's reactions in music listening with earable sensing and support music engagement-aware applications. We present three major design considerations to develop GrooveMeter.

\textbf{Unobtrusive sensing:} To support music engagement-aware applications in real-life situations, it is important to track reactions to music without relying on neither infrastructure-deployed nor excessive on-body sensors. 

\textbf{Accurate and robust detection:} While music listening reactions have distinctive semantic characteristics compared to daily human behavioral contexts, they sometimes show similar signal characteristics to each other. For example, a user's head nodding can be a reaction to a song being played, but it can also be a response that the user makes in face-to-face interaction with nearby people. Mumbling sounds can be made when a user hums along while listening to music or when he/she talks to him/herself. Since listening to music is often done as a secondary activity, such reaction-irrelevant behaviors, background noise, and other motion artifacts can degrade performance. It is critical to make reaction detection accurate and robust in daily music-listening situations.

\textbf{Low overhead:} Although GrooveMeter runs only while a user is listening to music, it is still important to have a low overhead. People already listen to music for quite a long time (2.63 hours daily in 2021~\cite{ml21}), and they would not prefer to consume a significant battery of their phone.

\subsection{GrooveMeter Architecture}

\begin{figure}[t]
	\centering
	\includegraphics[width=.7\textwidth]{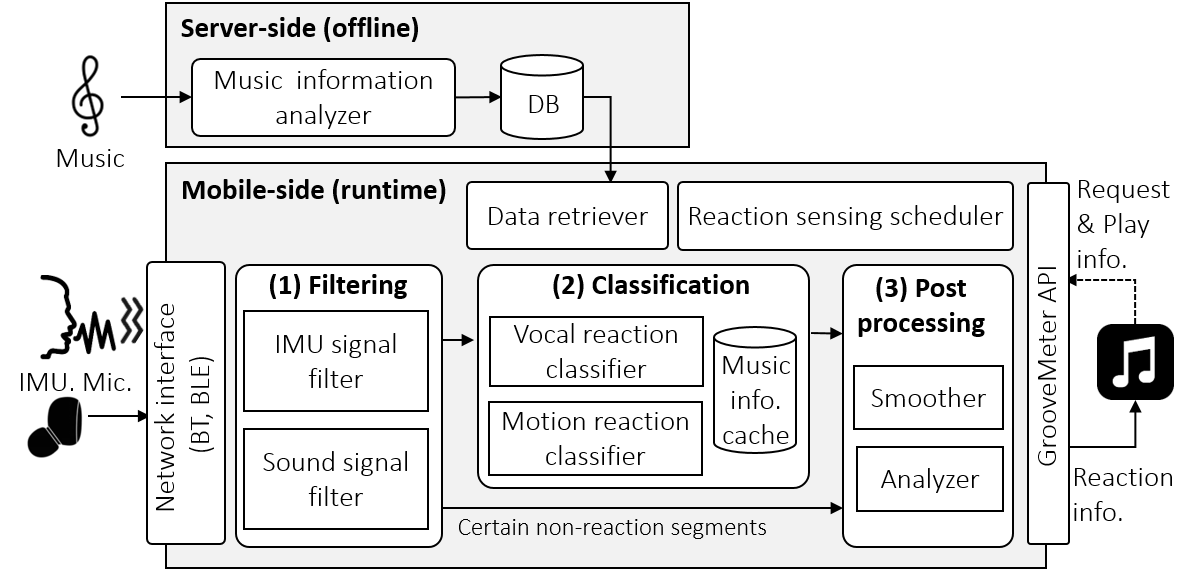}

	\caption{GrooveMeter architecture 
	\label{fig:system_flow}}
\vspace{-0.1in}
\end{figure}

Figure~\ref{fig:system_flow} shows the GrooveMeter architecture. It retrieves the music information of songs and maintains it in a database on the server. 
When a user starts to listen to a song, GrooveMeter activates audio and motion sensing on earbuds and sends the sensor stream to the smartphone; we currently use earbuds as merely a sensor stream provider due to their limited processing and programming capability. The whole processing is performed locally on the user's smartphone under the user's permission to avoid privacy concerns. The main operations of GrooveMeter are as follows.

\begin{enumerate}[leftmargin=*]
  \item As a first step, it investigates the signal characteristics and filters out data segments that can be certainly labelled as \textit{non-reaction} (\S\ref{subsec:vocal_filtering} and \S\ref{subsec:motion_filtering}).
  \item For uncertain segments, it identifies the reaction event from the classification model (\S\ref{subsec:vocal_classification} and \S\ref{subsec:motion_classfy}).
  \item GrooveMeter enhances the classification performance by leveraging music information retrieved from a song. GrooveMeter computes the similarity of sensory signals and music information, and corrects the label based on the similarity (\S\ref{subsec:vocal_correction}). 
  \item Based on the detected reaction events, GrooveMeter performs post-processing operations to provide high-fidelity information  
  (\S\ref{subsec:vocal_smoothing} and \S\ref{subsec:motion_post}). 
  GrooveMeter also adopts output smoothing in the post-processing operations of the vocal reaction detection (\S\ref{subsec:vocal_smoothing}).
\end{enumerate}

%% file: sections/04.system.tex
\section{Reaction Sensing}

%% file: sections/041.vocal.tex
\subsection{Vocal Reaction Detection~\label{subsec:vocal_reaction}}

\subsubsection{Challenges}
%\hfill

Here, we present several technical challenges that need to be addressed for the detection of vocal reactions.

\textbf{Vocal reaction events:} 
A straightforward way of detecting sound events like vocal reactions is to develop audio models or use pre-trained ones. However, developing a custom model requires a lot of effort for large-scale real-life data collection. Also, we find that simply adopting existing pre-trained models does not fit our purpose. Recently, a number of sound classification models have been released, which are pre-trained with an enormous amount of data and predict a large number of real-life sound events. One of the representative examples is YAMNet~\cite{yamnet}, released by Google in 2019. It is trained using more than 2 million YouTube videos and classifies 521 audio event classes. 

\begin{figure}[t]
    \centering
    \includegraphics[width=0.70\textwidth]{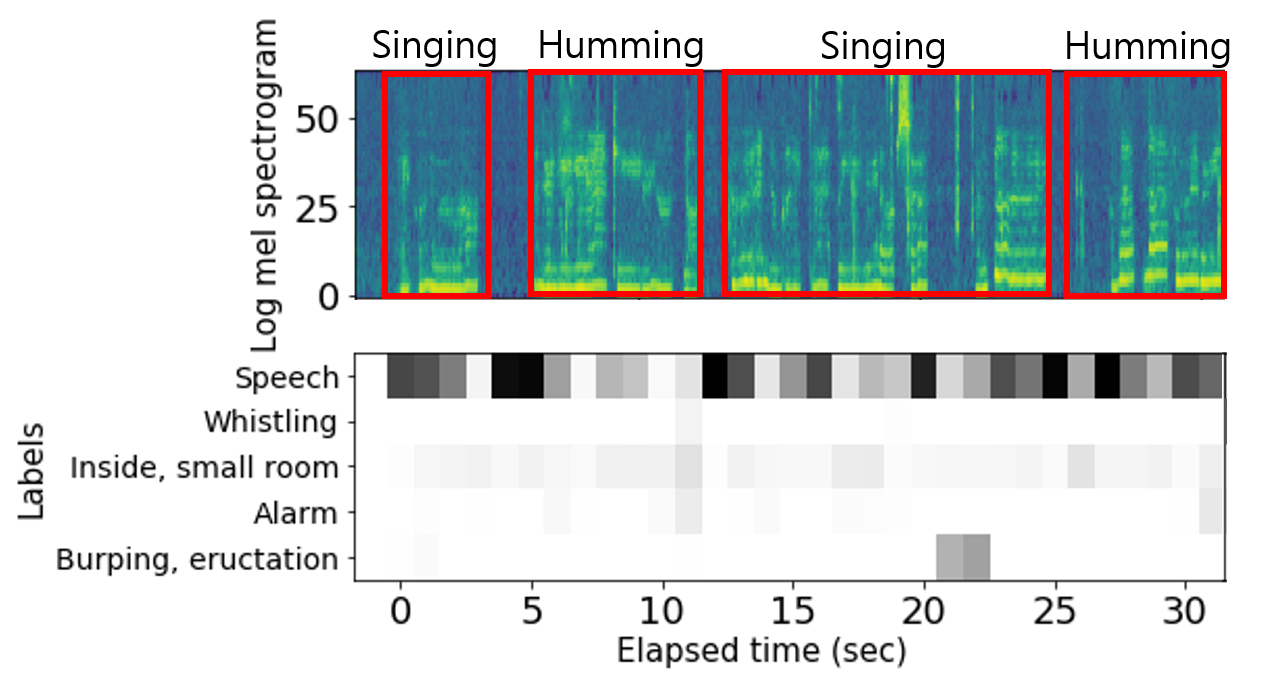}
    \vspace{-0.1in}
    \caption{YAMNet result for vocal reactions~\label{fig:yamnet_singing}}

    \vspace{-0.15in}
\end{figure}

While such pre-trained models show satisfactory performance for daily events, however, to the best of our knowledge, none of them includes our target events yet, i.e., vocal reactions in music listening, and accordingly they show poor performance on the reaction events. For example, YAMNet has the \textit{singing} label, but the corresponding audio data is mostly taken from video clips where a song is played with instruments, e.g., music video or a live clip of a band. However, when a listener makes a \textit{singing} reaction while listening to music via earbuds, the captured audio signal includes only the listener's singing voice without background music. Figure~\ref{fig:yamnet_singing} shows our preliminary study with YAMNet when we test its detection performance with an audio clip recorded while a listener sings along a song; the graph shows the output of the softmax layer. The result shows that YAMNet hardly selects the \textit{singing} label, but it classifies singing reactions mostly as the \textit{speech} label. 
The audio clips of \textit{whistling} are not labeled as speech but sometimes labeled as animal-related ones, e.g., birds singing. 
Developing a custom model with newly collected data could address these issues. However, it still requires collecting large-scale real-life sound events beyond reaction events to reflect user variability and avoid inference errors from various noises in real-life situations.

\begin{figure}[t]
\centering
    \includegraphics[width=0.70\textwidth]{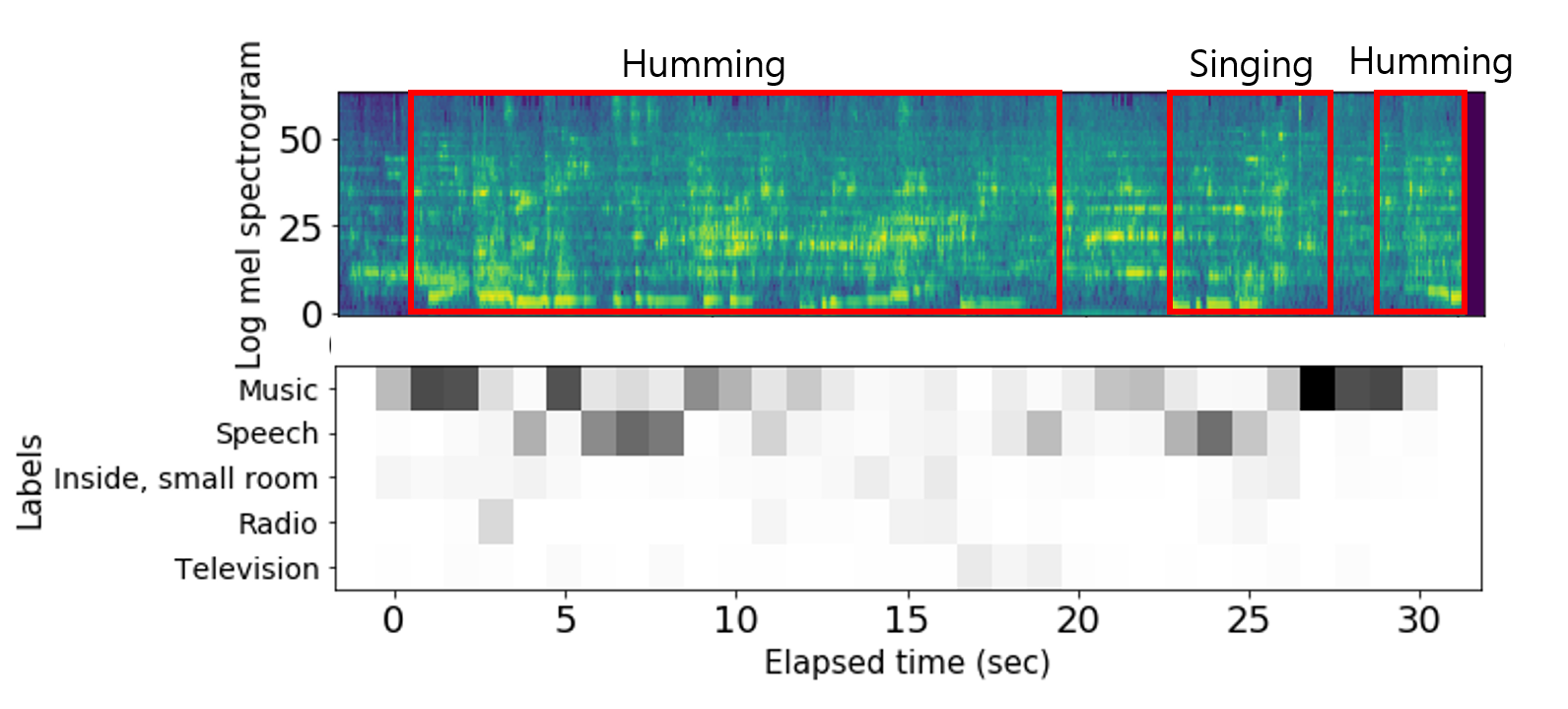} 
    \vspace{-0.1in}
    \caption{YAMNet result under background noise}
\label{fig:yamnet_reaction_with_noise}
\end{figure}

\textbf{Mixed with background noise:} Background noise (e.g., a sound of nearby people talking or background music in a cafe) makes it more complicated to detect vocal reactions correctly. We observe that YAMNet often classifies singing and humming reactions mixed with diverse noise in a cafe as the \textit{music} label. Figure~\ref{fig:yamnet_reaction_with_noise} shows such an example.  
Moreover, under a noisy condition, YAMNet's softmax scores of the reaction-relevant labels tend to be relatively low and sometimes even lower than reaction-irrelevant labels. From our data collected in noisy places such as a cafe and a car, only 5.3\% of singing/humming segments result in greater than 0.95 of YAMNet softmax score; in less noisy places such as lounge, 11.7\% does so. %\seungwoo{5.3\%: office, car, cafe, 11.7\%: lounge}

\textbf{Intermittence and alternation:} When a vocal reaction is made, it does not continue ceaselessly within a session, but sporadically and sometimes alternately with other reactions. For example, when a listener sings along, we observe intermittent, short-period pauses that the listener makes to breathe in the middle of the singing session. Also, listeners often make different types of vocal reactions alternately. For example, while listeners are singing along, they often switch to humming or whistling momentarily if they do not know the lyrics and come back to sing along again in the part where they know the lyrics.

\textbf{Processing cost:} Sound classification often involves processing-heavy operations such as MFCC computation and deep neural networks. While today's models provide optimization for on-device processing, e.g., MobileNet architecture of YAMNet, it still incurs significant processing and energy cost for continuous execution. 
For example, continuously performing audio classification with YAMNet on Galaxy S21 while playing songs incurs 3\% drop in battery level for one hour.

\begin{figure}[t]
    \centering
    \includegraphics[width=0.75\textwidth]{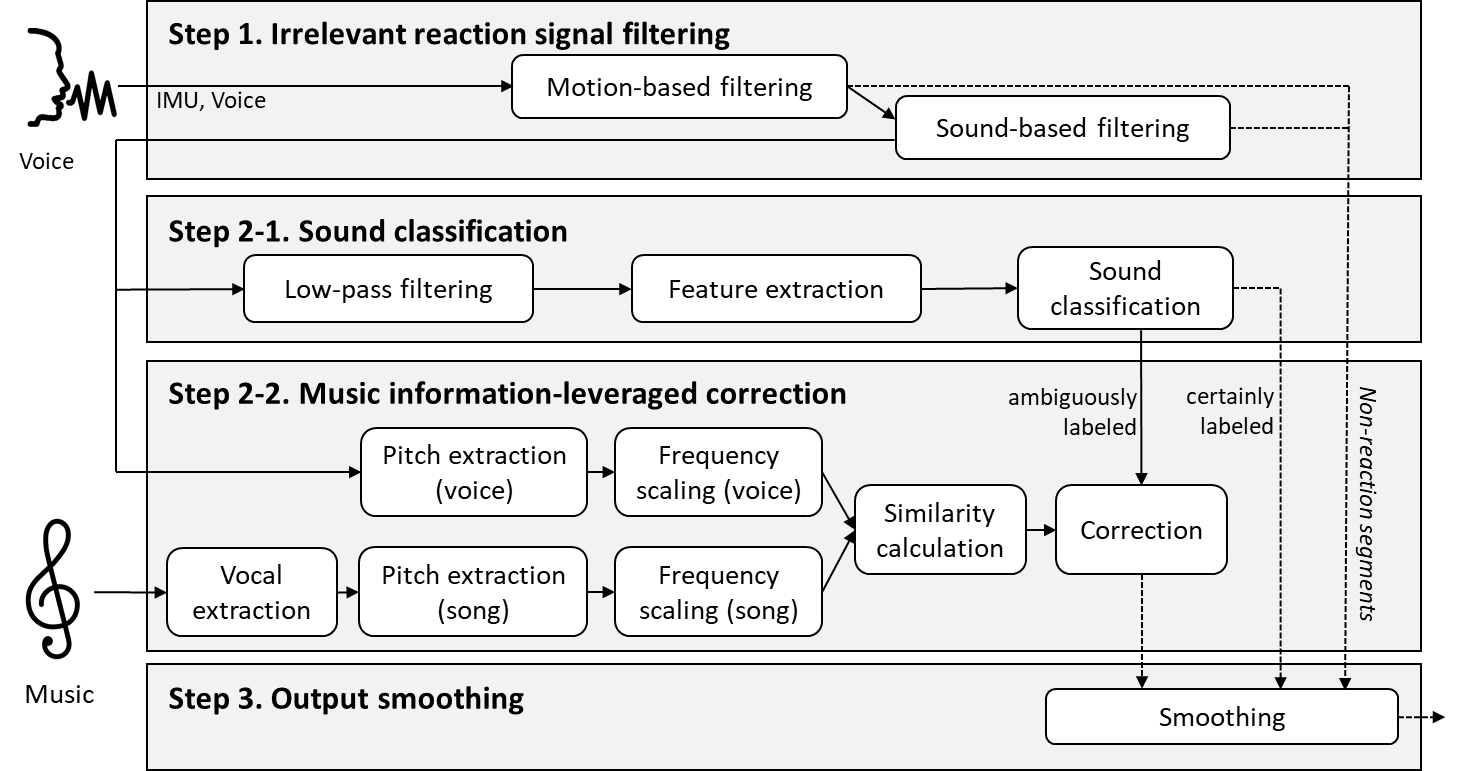}
    \caption{Vocal reaction detection pipeline }

    \label{fig:vocal_pipeline}

\vspace{-0.1in}
\end{figure}

\subsubsection{Overview}

To address the aforementioned challenges, we devise a novel pipeline that detects vocal reactions efficiently and reliably. Figure~\ref{fig:vocal_pipeline} shows its overview with three major operations. First, for cost saving, we adopt the early-stage filtering operation. Its key idea is to investigate the characteristic of audio and motion signals and filter out data segments that can be certainly labeled as \textit{non-reaction} (\S\ref{subsec:vocal_filtering}). Second, we initially classify sound events with the YAMNet model (\S\ref{subsec:vocal_classification}) and correct ambiguous labels by leveraging music information retrieved from a song being played (\S\ref{subsec:vocal_correction}). More specifically, it extracts the temporal pitch patterns from the audio signal and the song, and corrects the label by comparing the two. Last, it smooths the final outputs to cope with the momentarily introduced short, intermittent events and provides high-fidelity information (\S\ref{subsec:vocal_smoothing}.)

\subsubsection{Certain Non-reaction Signal Filtering~\label{subsec:vocal_filtering}}
%\hfill

\begin{figure}[t]
    \centering
    \mbox{
        \subfloat[Sound level\label{fig:vocal_sound_cdf}]{\includegraphics[width=0.32\columnwidth]{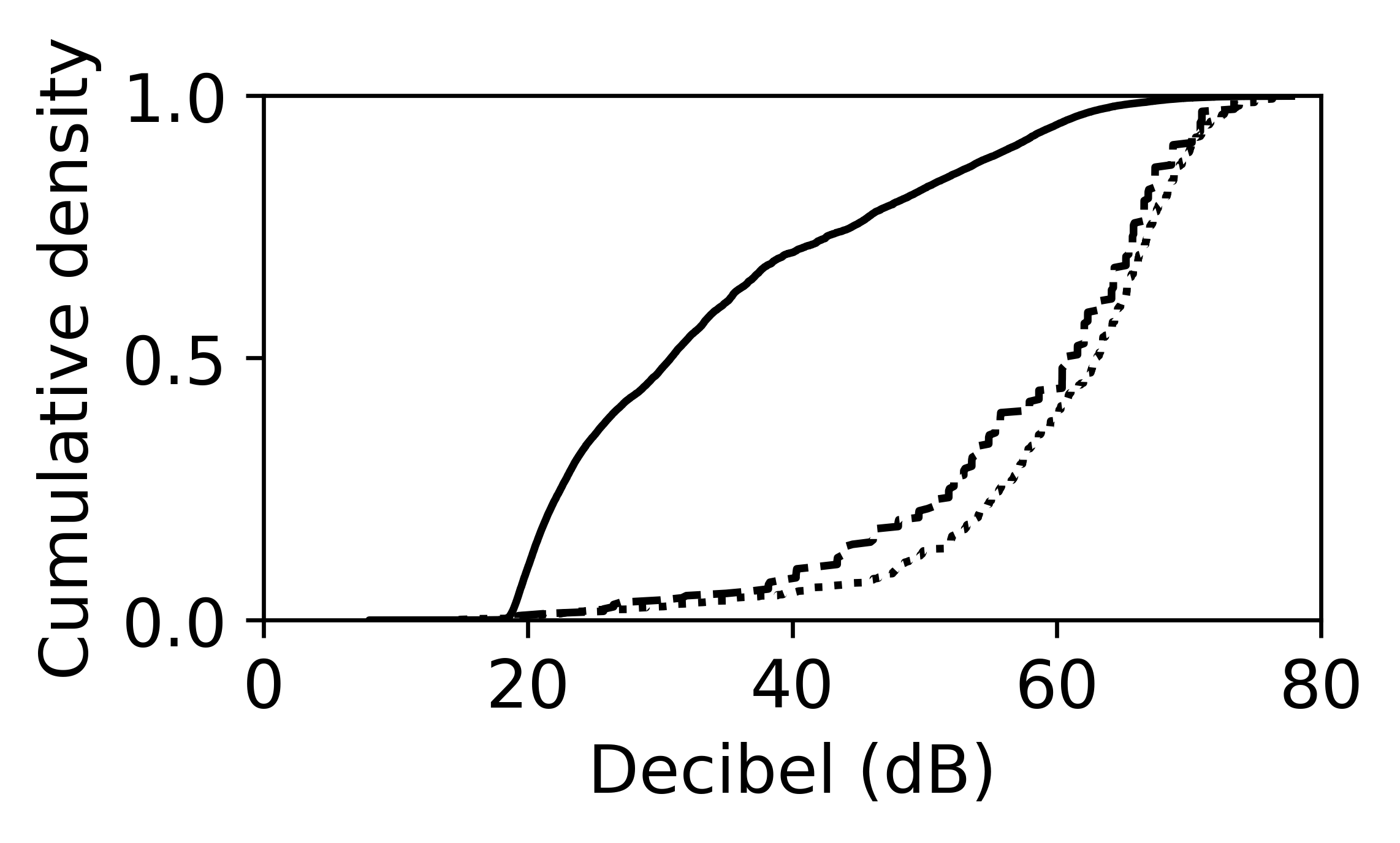}\vspace{-0.05in}}
        \hspace{0.2in}
        \subfloat[Movement level\label{fig:vocal_motion_cdf}]{\includegraphics[width=0.32\columnwidth]{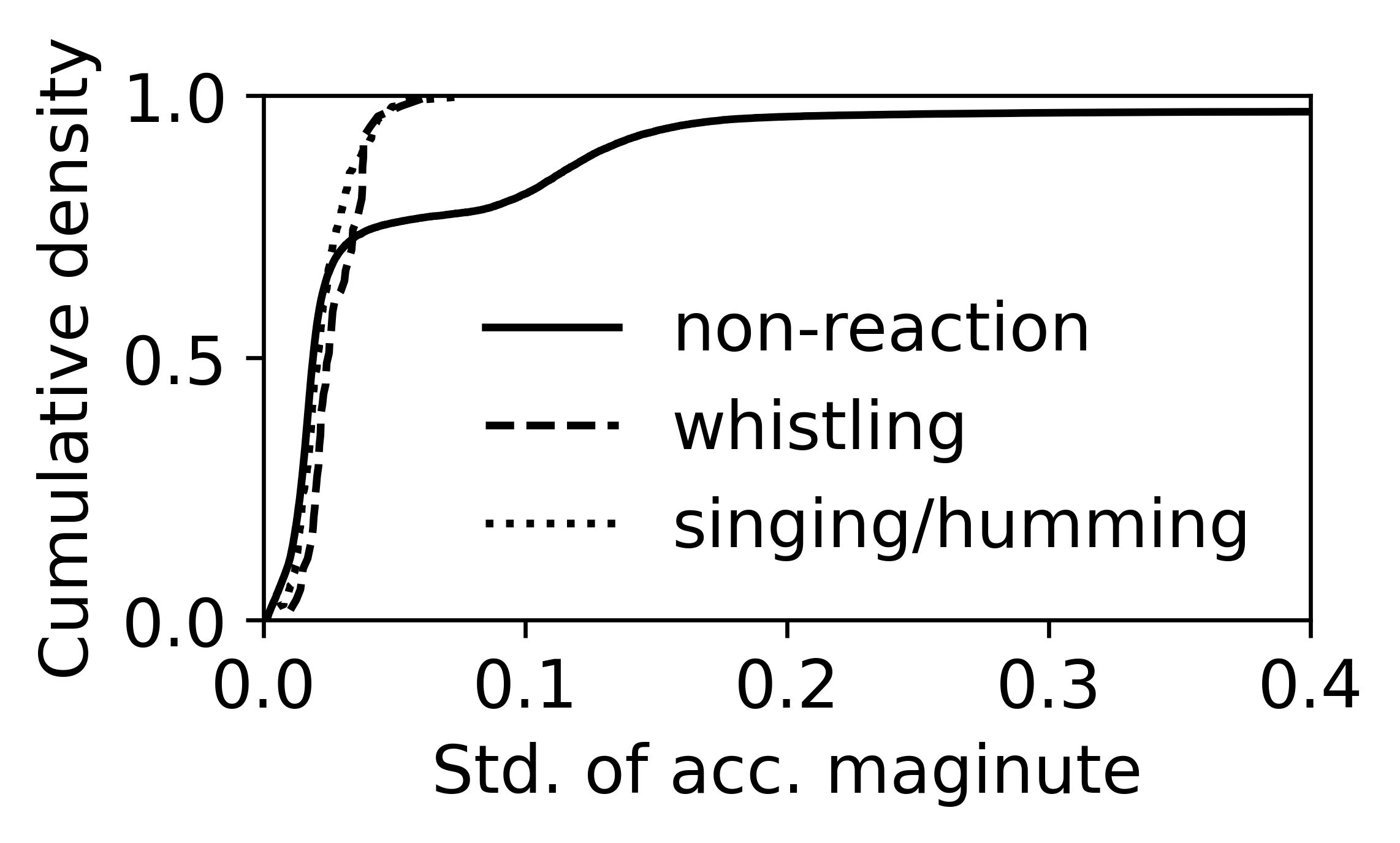}\vspace{-0.05in}}
        }
    \caption{CDF of sound and movement levels \label{fig:vocal_cdf}}

    \vspace{-0.15in}
\end{figure}

To save cost, we adopt an early-stage filtering operation. It identifies data segments that can be certainly labeled as \textit{non-reaction}, and avoids the processing-heavy classification operations for those segments. The identification logic is developed based on two observations. First, voice reactions would make sound events above a certain volume, especially due to a short distance between an audio source (an earbud wearer's mouth) and a microphone of an earbud. This implies that sounds below a certain volume threshold, e.g., silence and background noise, can be surely labeled as \textit{non-reaction}. Figure~\ref{fig:vocal_sound_cdf} shows the CDF of one-second decibel numbers for reaction and non-reaction sounds and validates our hypothesis. Second, more interestingly, voice reactions also incur a certain level of kinetic movement of an earbud. This is because a wearer's mouth movements to make vocal reactions activate the \textit{Zygomaticus} muscle located between a mouth and an ear, thereby triggering the impulse response in the inertial signals of an earbud. The recent works also showed that the earbud can capture the motion signal made by facial expression~\cite{lee2019automatic} and unvoiced sounds~\cite{khanna2021jawsense}. Similar to the first observation, it also implies that, if no motion is detected, the corresponding audio signal will be highly unlikely to belong to the \textit{reaction} label. Figure~\ref{fig:vocal_motion_cdf} shows the CDF of the standard deviation of one-second accelerometer magnitude. Interestingly, large motion is also associated with non-reaction label because listeners hardly make vocal reactions when they walk, run, and do exercise.

Based on such findings, we design a two-step filtering component. It first monitors the level of movement defined as the standard deviation of accelerometer magnitudes, and filters out data segments out of the threshold range (from the analysis of our dataset in \S\ref{sec:collection}, we set a range to 0.0104 and 0.12, respectively.) Second, for unfiltered segments, it monitors the decibel value of corresponding audio signals and filters out data segments if their decibel is lower than a threshold (in our implementation, 49 db). We put the motion-based filter in advance of the sound-based filter because the motion-based filter is more lightweight. From our measurement, the motion-based filter consumes 113 mW on Galaxy S21, while the sound-based filter does 134 mW.
It labels filtered segments as \textit{non-reaction} and delivers them to the post-processing operation without performing the subsequent operations.

\subsubsection{Sound Event Classification~\label{subsec:vocal_classification}}
%\hfill

\textbf{Target events:} \highlight{We target three types of vocal reaction events; \textit{singing (along)/humming}, \textit{whistling}, and \textit{non-reaction}. We combine singing and humming into the same class because they are often observed alternatively even in a single reaction session as mentioned above.}

\textbf{Preprocessing:} Audio data from earbuds are resampled at 16 kHz and divided into 1 second-long segment. We then apply a Low Pass Filter (LPF) of order 1 (cut-off at 2 kHz) on the data segments to reduce noise signals with frequencies higher than the major frequency range of our target vocal reactions. 

\textbf{Audio feature extraction:} The preprocessed segments are then converted to spectrograms using the Short-Time Fourier Transform with a periodic Hann window; we set a window size and a window hop to 25 ms and 10 ms, respectively. Then, we map the spectrogram to 64 mel bins with the range between 125 and 7,500 Hz and compute log mel spectrograms by applying log. Finally, we frame the features into a matrix of $96\times64$ (96 frames of 10 ms each and 64 mel bands of each frame.

\begin{table}[]
%\footnotesize
\centering
\caption{Mapping from YAMNet to GrooveMeter labels}
\label{table:vocal_mapping}

\begin{tabular}{ll}
 \hline
 \textbf{YAMNet} & \textbf{GrooveMeter} \\ \hline
 humming, singing & singing/humming \\ 
 whistling, whistle & whistling \\ 
 speech, music & ambiguous (candidate for singing/humming or non-reaction) \\ 
 others & non-reaction \\ \hline
\end{tabular}

\end{table}

\textbf{Classification and labeling:} We use YAMNet~\cite{yamnet} as a base component for the sound event classification. However, since the taxonomy of YAMNet labels does not well fit our target classes, we construct a mapping from YAMNet to GrooveMeter labels as shown in Table~\ref{table:vocal_mapping}. As discussed before, YAMNet is poor in discriminating \textit{singing} reactions from speech and music signals as shown in Figures~\ref{fig:yamnet_singing} and \ref{fig:yamnet_reaction_with_noise}. We thus map speech and music labels from YAMNet classification output to \textit{ambiguous}. We perform further investigation for ambiguous segments with the subsequent correction operation. Other labels are directly sent to post processing operations.

\textbf{Rank constraint relaxation:} Simply relying on the YAMNet's classification result is not sufficient for accurate vocal reaction detection even with applying label mapping. Due to the background noise and characteristics of vocal reactions, YANNet outputs speech or music as a top-1 classification label only for 57.8\% of singing/humming segments in our dataset.
The ratio increases as we include speech or music label in a lower rank, i.e., 65.4\% (top 2), 70.1\% (top 3), 73.3\% (top 4), and 75.7\% (top 5), etc. The whistling segments also show a similar characteristic.

To address the problem, we employ a rank constraint relaxation policy, allowing some of segments that YAMNet does not classify as one of our target labels to go through the correction step. Here we need a balance to avoid unnecessary cost for the correction step and increase of false positive errors due to additional non-reaction segments that can be incorrectly classified as vocal reaction labels. We do not apply relaxation if the quality of the classification output by YAMNet is highly good enough. Otherwise, we check whether lower ranked outputs include some of vocal reaction labels. If so, we consider it as a \textit{uncertain} label that needs further investigation with the correction step. 

To quantify the quality of the classification output, we adopt the strategy of uncertainty sampling proposed in the domain of active learning. Several techniques have been developed to calculate an inference instance’s uncertainty, such as least margin~\cite{scheffer2001active} and highest entropy~~\cite{shannon1948mathematical}. Currently, we choose the least margin, which measures the uncertainty by taking the difference between the confidence values of the top two output classes. If the margin of a segment is lower than a threshold, we check top-k YAMNet output labels. If they include our target labels, i.e., speech, music, humming, singing, whistling, and whistle, the segment is considered uncertainly labeled, e.g., uncertain humming, and it is forwarded to the correction step.  
In our current implementation, we empirically set the margin threshold and k to 0.9 and 5, respectively.

\subsubsection{Music information-leveraged Correction~\label{subsec:vocal_correction}}
%\hfill

We finalize ambiguous or uncertain segments from the previous step by leveraging music information of a song being played. Based on the prosodic similarity between the audio signal (from earbud) and the song, we correct ambiguous segments with speech or music labels to \textit{singing/humming} or \textit{non-reaction}. Also, we deal with uncertain segments in the same way.

\begin{figure*}
    \centering
    \begin{subfigure}[b]{0.35\textwidth}
        \centering
        \includegraphics[width=\textwidth]{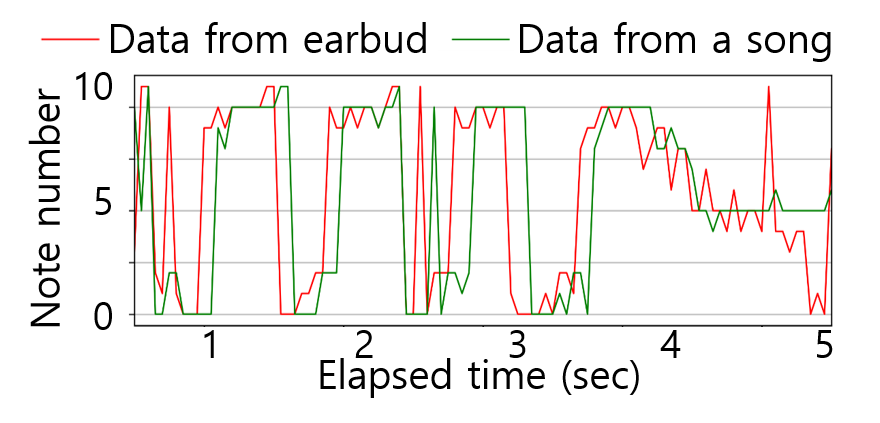}
        \caption[]%
        {{\small Singing along}}    
        \label{fig:singing_note}
    \end{subfigure}
    \hspace{0.2in}
    \begin{subfigure}[b]{0.35\textwidth}  
        \centering 
        \includegraphics[width=\textwidth]{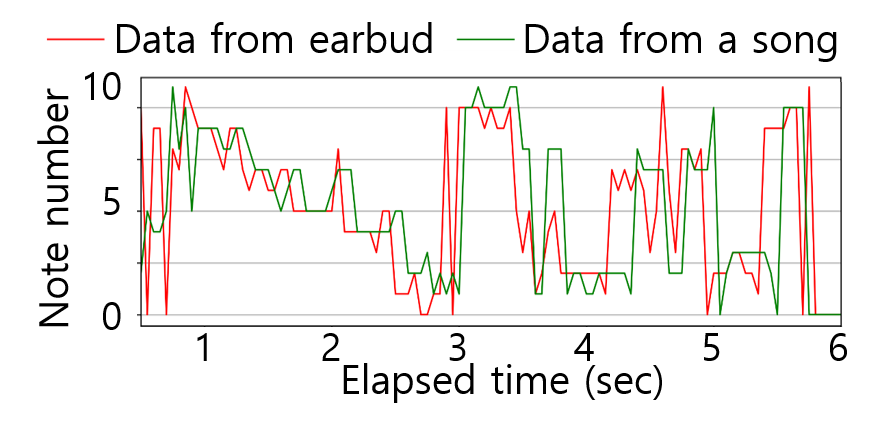}
        \caption[]%
        {{\small Humming}}    
        \label{fig:humming_note}
    \end{subfigure}
    \vskip\baselineskip
    \begin{subfigure}[b]{0.35\textwidth}   
        \centering 
        \includegraphics[width=\textwidth]{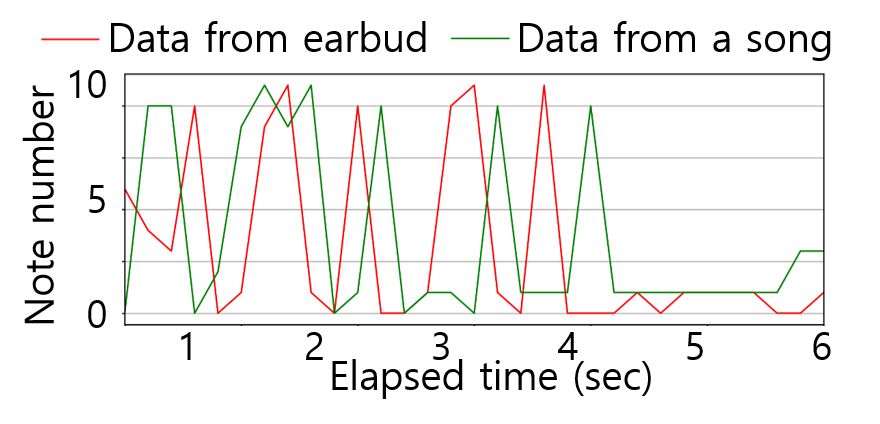}
        \caption[]%
        {{\small Whistling}}    
        \label{fig:whistling_note}
    \end{subfigure}
    \hspace{0.2in}
    \begin{subfigure}[b]{0.35\textwidth}   
        \centering 
        \includegraphics[width=\textwidth]{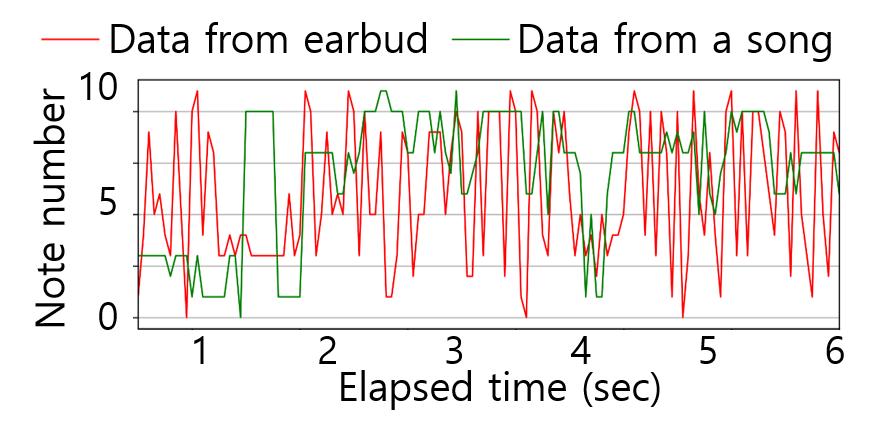}
        \caption[]%
        {{\small Non-reaction}}    
        \label{fig:non_note}
    \end{subfigure}
    \vspace{-0.1in}
    \caption[]
    {\small Notes of a chromatic scale} 
    \label{fig:notes}
    \vspace{-0.15in}
\end{figure*}

\textbf{Prosodic similarity computation:} 
To measure the prosodic similarity between a vocal signal and a song, we consider the melody which refers to a linear succession of musical tones. Our key intuition is that vocal reactions would follow the sequence of notes of a song being played, but reaction-irrelevant speech signal would not. Figure~\ref{fig:vocal_pipeline} shows the detailed procedure (See Step 2-2.) To extract the sequence of a note, we first extract the pitch information using CREPE~\cite{kim2018crepe}, a state-of-the-art pitch tracker; pitch, i.e., frequency information, is extracted every 0.1 seconds. Then, we convert the pitch (frequency) to a musical note with an octave number. We convert again it to a 12-tone chromatic scale, i.e., without octave number, because we observe that vocal reactions are often played an octave higher or lower than a song being played. 
For the audio file of a song, we perform the vocal extraction prior to the pitch extraction to focus on the predominant melodic line of music. This insight also comes from our observation that vocal reactions would mostly follow vocals (singing voice) rather than instruments. We use Spleeter~\cite{spleeter2020} to separate a vocal source from a song.

\begin{figure}[t]
    \centering
    \mbox{
        \subfloat[Lounge\label{fig:rest_similarity}]{\includegraphics[width=0.25\columnwidth]{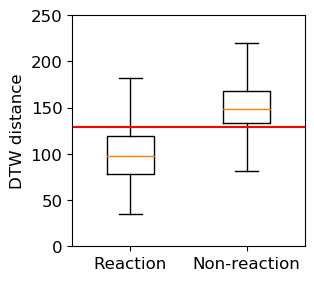}\vspace{-0.05in}\hspace{-0.02in}}
        \subfloat[Office\label{fig:office_similarity}]{\includegraphics[width=0.25\columnwidth]{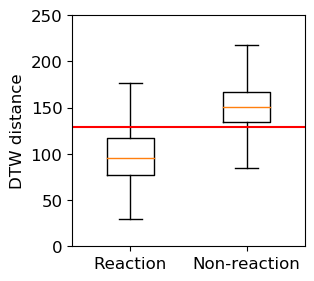}\vspace{-0.05in}\hspace{-0.02in}}
        \subfloat[Car\label{fig:car_similarity}]{\includegraphics[width=0.25\columnwidth]{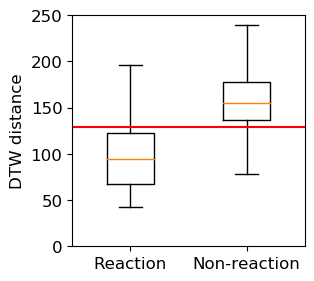}\vspace{-0.05in}\hspace{-0.02in}}
        \subfloat[Cafe\label{fig:cafe_example}]{\includegraphics[width=0.25\columnwidth]{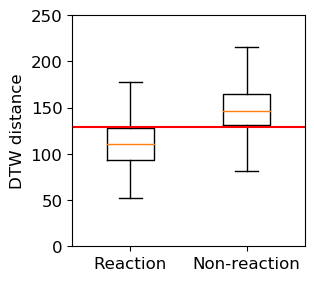}\vspace{-0.05in}\hspace{-0.02in}}
        
        }
    \vspace{-0.1in}
    \caption{Similarity differences 
    \label{fig:similarity_distribution}}
    
\end{figure}

We compute the similarity between two sequences of notes (one from a user's vocal signal and the other from a song) and make a final decision. We map 12 notes (C, C\#, D, D\#, E, F, F\#, G, G\#, A, A\#, B) to twelve integer values (0 to 11). 
Figure~\ref{fig:singing_note}, \ref{fig:humming_note}, and \ref{fig:whistling_note} show a high correlation of note patterns with the playing song, the non-reaction part does not (Figure~\ref{fig:non_note}). We consider dynamic time warping (DTW) as a similarity measure function because two note patterns can vary in speed. 

We label a segment as \textit{non-reaction} if the DTW distance is larger than a threshold. Otherwise, we apply label mapping and confirm its final label, i.e., speech and music to \textit{singing/humming}, whistling and whistle to \textit{whistling}, humming and singing to \textit{singing/humming}. In our implementation, we empirically set the threshold to 130.
Figure~\ref{fig:similarity_distribution} presents the distribution of DTW distance values of reaction (i.e., singing/humming) and non-reaction segments in our dataset. Interestingly, the distance shows similar distribution regardless of noise conditions. The median distance of reaction segments in the cafe case, which is the most noisy condition in our dataset, is slightly larger than that of other cases. Still, most of reaction segments can be distinguished from non-reaction segments.

\subsubsection{Post processing~\label{subsec:vocal_smoothing}}
%\hfill

\textbf{Smoothing:} We use a Hidden Markov Model (HMM) to smooth the classification output. The key idea is to train the HMM model from the sequence of the classification outputs of the training dataset and to use the trained model for the output smoothing. We define the observation sequence as a sequence of the classification outputs and perform smoothing by estimating the optimal sequence of hidden states, which can be mapped to the smoothed sequence of reaction events. More specifically, for a given sequence of classification outputs at time $t$, $O^{t}$ = ( $o_{1}$, … $o_{t}$ ), we extract the sequence of hidden states with the maximum probability, $S^{t}$ = ( $s_{1}$, … $s_{t-1}$ ), from time $1$ to $t-1$. Then, the smoothed value at time $t$, ${\hat{s}}_{t}$, is obtained as follows:
\begin{equation}\label{eq1} 
\hat{s}_{t}=\underset{s_{t}}{\mathrm{argmax}}p(s_{t}|O^{t+1},\lambda) 
\end{equation} 
We apply the Viterbi algorithm~\cite{forney1973viterbi} for efficient computation of maximum probability and use the 6 second-long window as an input sequence, i.e., a sequence of recent 6 classification outputs. The smoothing operation can be omitted if an application prefers the real-time output for interactive service.

\textbf{High-fidelity information:} The primitive information GrooveMeter provides is the occurrence of vocal reaction events, i.e., reaction type and start and end time. We can further provide high-fidelity information by aggregating events, e.g., which verse was sung along the most, and by analyzing audio signals, e.g., scoring vocal reactions based on the prosodic similarity. Defining such information and assessing its usability will be needed to be considered differently depending on the application requirement. We leave it for future work.

%% file: sections/042.motion.tex
\subsection{Motion Reaction Detection~\label{subsec:motion_reaction}}

\subsubsection{Challenges}
%\hfill

We present challenges we considered to design the pipeline for motion reaction detection.

\begin{figure*}
    \centering
    \begin{subfigure}[b]{0.4\textwidth}
        \centering
        \includegraphics[width=\textwidth]{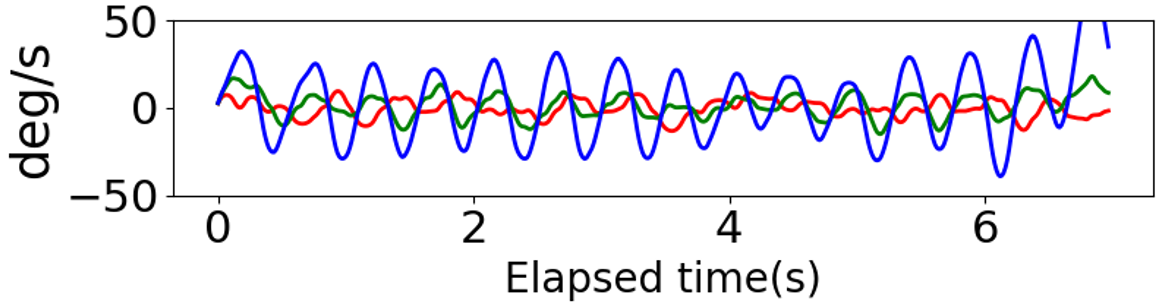}
        \caption[]%
        {{\small Head nodding - case 1}}    
        \label{fig:nodding_big}
    \end{subfigure}
    \hspace{0.47in}
    \begin{subfigure}[b]{0.4\textwidth}  
        \centering 
        \includegraphics[width=\textwidth]{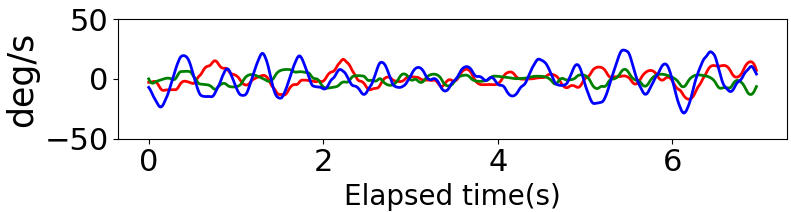}
        \caption[]%
        {{\small Head nodding - case 2}}    
        \label{fig:nodding_small}
    \end{subfigure}
    \vskip\baselineskip
    \begin{subfigure}[b]{0.4\textwidth}   
        \centering 
        \includegraphics[width=\textwidth]{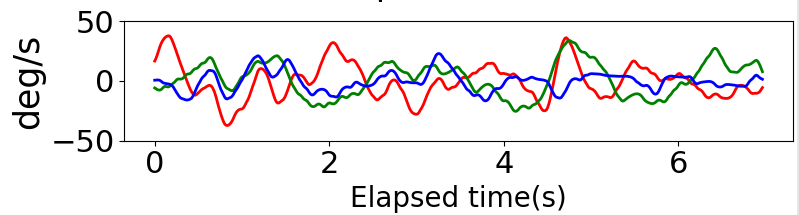}
        \caption[]%
        {{\small Head motion along with body swaying - case 1}}    
        \label{fig:body_large}
    \end{subfigure}
    \hspace{0.47in}
    \begin{subfigure}[b]{0.4\textwidth}   
        \centering 
        \includegraphics[width=\textwidth]{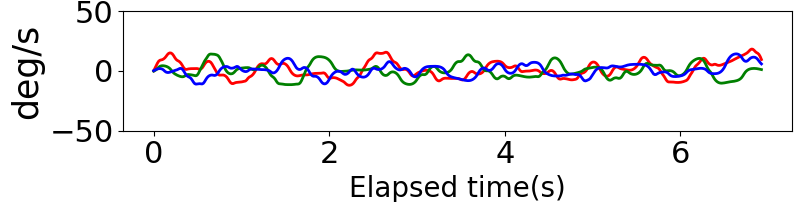}
        \caption[]%
        {{\small Head motion along with body swaying - case 2}}    
        \label{fig:body_small}
    \end{subfigure}
    \caption[]
    {\small Diverse patterns of motion reaction} 
    \label{fig:motion_reaction_samples}
\end{figure*}

\textbf{Periodic and repetitive, but diverse motion trajectories:} From our observation, we found out that motion reaction often exhibits repetitive, periodic patterns. \highlight{For example, head nodding (Figure~\ref{fig:nodding_big} and \ref{fig:nodding_small}) continues for a certain duration of time with a regular pattern, which yields a signal waveform with a certain level of periodicity.} One may argue that such repetitive patterns can be easily captured by typical pipelines for physical activity recognition. \highlight{However, we found out that classifiers using widely-used features representing the signal's periodicity or statistical features were not effective for in-the-wild reaction data as shown in \S\ref{subsec:motion_performance}.} This is mainly because, unlike well-defined activities or gestures following typical motion trajectories, the movement behavior of real-life motion reactions tends to vary. For example, people often move their head up and down while listening to music, but sometimes they move or tilt from side to side. The magnitude and speed of motion reaction also vary even for the same user or the same listening session. \highlight{Figure~\ref{fig:nodding_big} and \ref{fig:nodding_small} show two different patterns of head nodding from the same participant and Figure~\ref{fig:body_large} and \ref{fig:body_small} show head motion along with body swaying from another participant.} %\seungwoo{invalid participant ID}

\begin{figure}[t]
    \centering
    \mbox{
        \subfloat[Riding in a car\label{fig:motion_non_car_1}]{\includegraphics[width=0.4\textwidth]{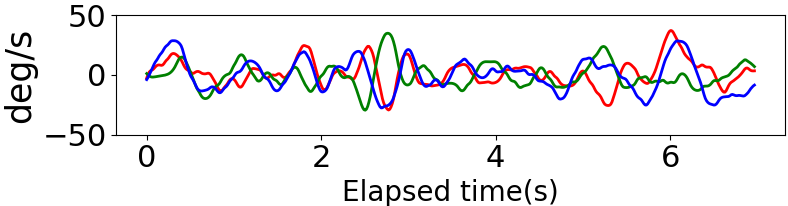}}
        \hspace{0.47in}
        \subfloat[Chewing a cookie\label{fig:motion_non_cookie_1}]{\includegraphics[width=0.4\textwidth]{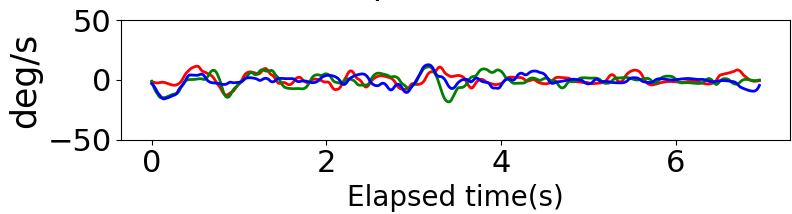}}

        }
    
    \caption{Repetitive, reaction-irrelevant motions~\label{fig:motion_non_reaction_samples}}
    
\end{figure}

\textbf{Confusing motion trajectories from reaction-irrelevant movements:} According to our preliminary study with in-the-wild data, we found out that there are several reaction-irrelevant movements that show repetitive signal patterns and accordingly can cause classification errors in reaction detection. For example, when a person is riding in a car, the movement and vibration of the car can cause the repetitive IMU signal on earbuds, even though the person does not make any other explicit motion (see Figure~\ref{fig:motion_non_car_1}). Similarly, Figure~\ref{fig:motion_non_cookie_1} shows that chewing a cookie also generates a certain level of repetitive patterns.

\begin{figure}[t]
    
	\centering
	\includegraphics[width=0.4\textwidth]{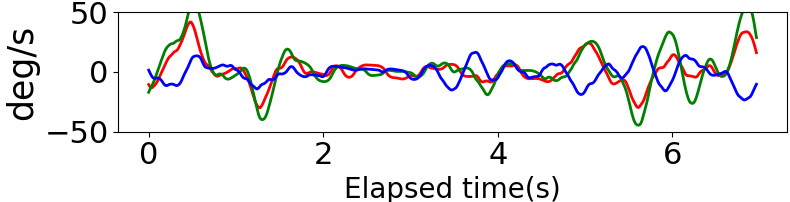}
	
	\caption{Nodding while performing an other activity~ \label{fig:nodding_body}}
	
\end{figure}

\textbf{Affected by other motion artifacts:} People often listen to music as a secondary activity, which means that their primary task (e.g., working at an office, relaxing at a cafe) and corresponding movement (e.g., typing keyboard, chewing a cookie) can affect the IMU signal. Figure~\ref{fig:nodding_body} shows the example of the gyroscope data when a user is nodding to the rhythm while performing another activity. As seen, the periodicity from the nodding movement is less clearly shown, compared to Figure~\ref{fig:nodding_big} and Figure~\ref{fig:nodding_small}.

\subsubsection{Overview}

\begin{figure}[t!]
    \centering
    \includegraphics[width=0.75\columnwidth]{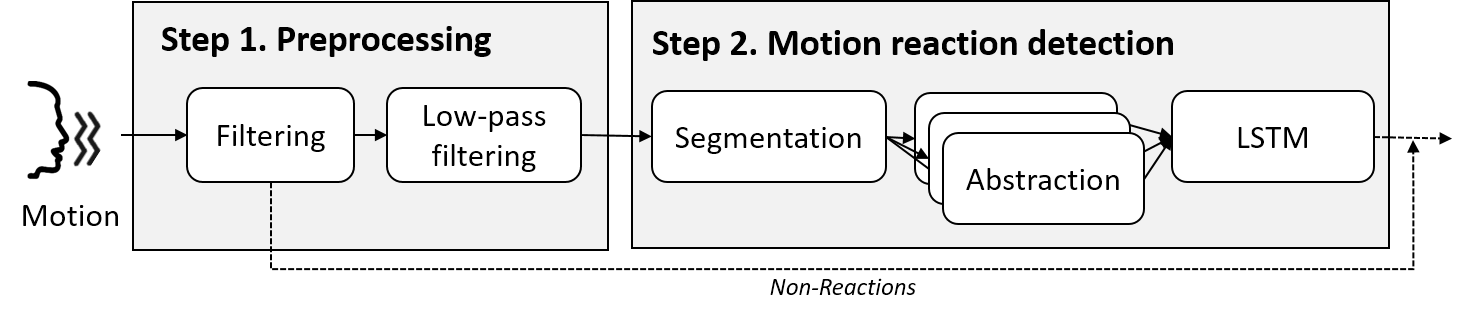}
    
    \caption{Motion reaction detection pipeline.\label{fig:motion_pipeline}}

\end{figure}

We design a novel pipeline for motion reaction detection that addresses the aforementioned challenges. Figure~\ref{fig:motion_pipeline} shows its overview with two main operations, preprocessing and reaction detection. In the preprocessing stage, we adopt a simple filter to avoid unnecessary processing for the classification operation. It early determines the output label of input data that are not highly likely reaction, e.g., no movement or too large movement, as \textit{non-reaction} (\S\ref{subsec:motion_filtering}). Then, we remove the noise caused by other motion artifacts by adopting a low-pass filter (\S\ref{subsec:motion_noise_removal}). Second, we abstract raw IMU signals and extract a sequence of \textit{motion units} to represent an abstraction of a user's motion pattern. With the sequence, we detect motion reaction using LSTM, \highlight{performing binary classification (head motion vs. non-reaction)} (\S\ref{subsec:motion_classfy}).

\subsubsection{Reaction-irrelevant movement filtering~\label{subsec:motion_filtering}}
%\hfill

We design a simple threshold-based filter based on our observation. It sorts out reaction-irrelevant data by looking into the movement level. We define the movement level as the standard deviation of a one-second segment of accelerometer signal. 
Similarly in the motion filter of vocal reaction detection (\S\ref{subsec:vocal_filtering}), it is obvious that no movement implies \textit{non-reaction}. Large movement is also associated with a \textit{non-reaction} label, because it is very unlikely that listeners nod their head while doing workout, walking, or running.

\begin{figure}[]
\begin{minipage}[t]{0.35\linewidth}
    \includegraphics[width=\columnwidth]{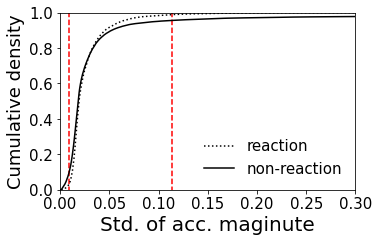}
    
    \caption{CDF of movement level of motion reactions. \label{fig:motion_threshold}}
    
\end{minipage}%
    \hspace{0.3in}
\begin{minipage}[t]{0.25\linewidth}
    \includegraphics[width=\columnwidth]{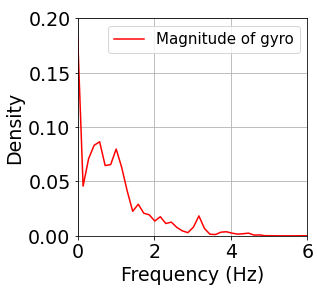}
    
    \caption{Dominant frequency distribution of reaction data.}

    \label{fig:dominant_freq}
\end{minipage} 
\vspace{-0.1in}
\end{figure}

For the implementation, we carefully select a threshold range that allows us to sort out as many non-reaction cases as possible without missing reaction cases. To this end, we investigate the movement level of our dataset. As shown in Figure~\ref{fig:motion_threshold}, we examine the CDF of the movement level and empirically set the low and high threshold values to 0.0092 g and 0.114 g, respectively.

\subsubsection{Noise removal~\label{subsec:motion_noise_removal}} 
The next step is to remove motion noise caused by other motion artifacts. The intuition behind this idea is that a listener’s motion reaction tends to follow beat patterns of a song. Accordingly, we could expect that motion reactions tend to exhibit low frequency movement, considering that typical tempo of common music genres ranges between 60 and 180 beats per minutes. Figure~\ref{fig:dominant_freq} shows a distribution of dominant frequency extracted from our motion reaction data. Almost all dominant frequencies are less than 4 Hz. 
%Also, we observe that the motion artifacts due to vibration from a moving car induces relatively high frequency components \needtofill{(see Figure~\ref{})}. 
Thus, we process the raw IMU data sampled at 70 Hz with a low pass filter (LPF) of order 1 (cut-off at 5 Hz), allowing some margins.

\subsubsection{Motion reaction classification~\label{subsec:motion_classfy}}

The classification operation includes two major steps. The first step is to extract the sequence of \textit{motion abstraction}. With IMU data, we extract temporal motion patterns. As mentioned before, motion reactions do not show well-defined, typical movement trajectory and duration. We devise a method that abstracts IMU signals reflecting human motions in a simple way and detects motion reactions accordingly. For abstraction, we define a \textit{motion unit} that represents a set of features derived from a short time interval of IMU data. We extract motion units from IMU signals and yield a sequence of motion units. The sequence can be viewed as an abstraction of the user’s motion pattern. It can represent motion reactions to music, random body motion irrelevant to the reactions, or stationary state. To extract motion units, we segment the IMU data processed by the LPF into the length of 100ms. For the segment, we compute a set of statistical features for each axis of gyroscope, i.e., max, min, mean, range, standard deviation, and RMS.

The next step is to classify a sequence of motion units into one of two classes, \highlight{head motion and non-reaction}. Considering that the input is a temporal sequence, we adopt an LSTM model, widely used for the prediction of sequential data such as time series. We choose the LSTM based on our experiment showing better performance compared with other methods such as RNN and GRU.
We build a classification model consisting of an LSTM layer with 32 hidden units, a dropout layer with a drop rate of 0.5, a ReLU layer, and a softmax layer. The model is implemented using Keras API and trained with up to 300 epochs using an Adam optimizer. We empirically set a window size for the classification to 7 seconds, where the $F_1$ score starts to saturate. A window of data is framed into a matrix of $70\times 18$ to be fed into the model.

\subsubsection{Post processing~\label{subsec:motion_post}}
GrooveMeter combines classifier outputs and filtered results as non-reaction to provide final inference output. The primitive information GrooveMeter provides is the occurrence of motion reaction events. It further provides detailed analytics by aggregating event information, e.g., the part of a song for which listeners move the most, songs that make listeners move frequently.

%% file: sections/06.datacollection.tex
\section{Music Reaction Data~\label{sec:collection}}

To build and evaluate GrooveMeter, we create MusicReactionSet, a novel dataset consisting of audio and IMU data from a variety of music listening reactions. To the best of our knowledge, this is the first dataset targeting reactions in music listening situations. The data collection was conducted under the IRB approval.

\textbf{Participants:} We recruited 30 participants (M: 18, F: 12) from a university campus, $P_A1$ to $P_A30$; their ages were between 20-26 (mean 22.7). We obtained informed consent from the participants. All of them reported that they frequently listen to music in daily life. After completing data collection, they were compensated with a gift card worth USD 18.

\begin{table}[]
\footnotesize
\centering
\caption{Characteristics of music listening situations }
\label{table:data_situations}
\begin{tabular}{lcc}
 \hline
 \textbf{Situation} & \textbf{Reaction-irrelevant motions} & \textbf{Background Noise} \\ \hline
 \thead{Resting in a lounge} & \makecell{random movement while sitting on a chair} & \makecell{noise from air-purifier,\\ murmuring sound outside, ...}\\ 
 \thead{Working at an office} & \makecell{motions during web search and word processing} & \makecell{keyboard typing,\\ mouse clicking sound, ...}\\ 
 \thead{Riding in a car} & bouncing along the road & various noises of driving car \\ 
 \thead{Relaxing at a cafe} & \thead{drinking coffee,chewing a cookie} & \makecell{background music,\\ nearby conversation, chewing sound, ...} \\ \hline
\end{tabular}
\end{table}

\begin{figure}[t]
    \centering
    \mbox{
        \subfloat[Lounge\label{fig:lounge}]{\includegraphics[width=0.22\columnwidth]{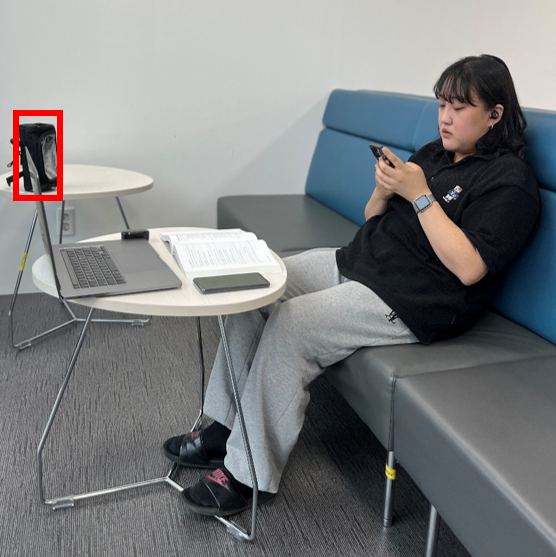}\hspace{0.07in}}
        \subfloat[Office\label{fig:office}]{\includegraphics[width=0.22\columnwidth]{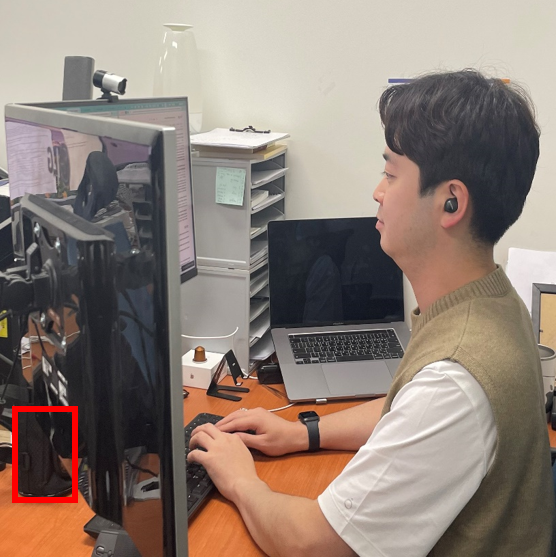}\hspace{0.07in}}
        \subfloat[Car\label{fig:car}]{\includegraphics[width=0.22\columnwidth]{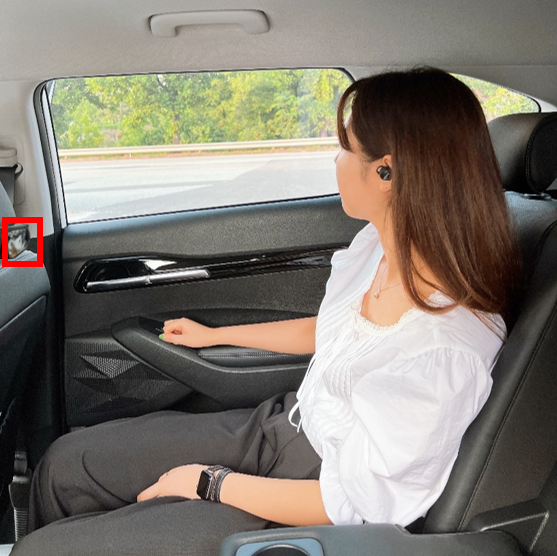}\hspace{0.07in}}
        \subfloat[Cafe\label{fig:cafe}]{\includegraphics[width=0.22\columnwidth]{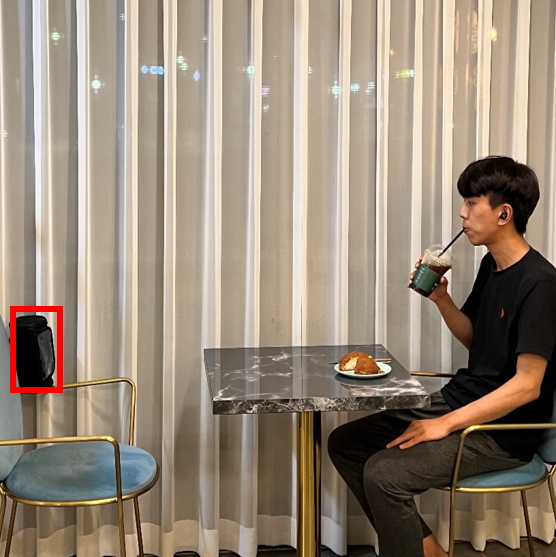}\hspace{0.07in}}
        }
    
    \vspace{-0.1in}
    \caption{In-the-wild data collection in various places. \label{fig:data_collection}}
    
    \vspace{-0.1in}
\end{figure}

\textbf{In-the-wild data collection:} Each participant was invited to four places and asked to listen to a set of songs while doing other activities, i.e., resting in a lounge, working at an office, riding in a car, and relaxing at a cafe (see Figure~\ref{fig:data_collection}). We consider these places to (a) reflect diverse real-life situations where people often enjoy listening to music and (b) investigate the impact of diverse audio and motion noise on reaction detection. Table~\ref{table:data_situations} shows the characteristics of four situations. Note that the car riding case reflects other similar situations, e.g., in public transportation such as taxi and bus. 

More specifically, in each situation, the participants freely chose three songs with two different genres, i.e., exciting/up-tempo and slow/soft, from the top-50 chart in a music streaming service. They listened to the songs using earbud devices and a smartphone provided. To collect data from their natural reactions, we did not give any instruction about making reactions and also let them be alone in the places (except riding in a car). Note that we did not include the data from the first song because the participants often felt a little distracted right after they moved to a new place. 
Finally, 926 minute-long IMU and audio data were collected in total from 240 music listening sessions (8 sessions per participant). Half of them are exciting/up-tempo songs, while the rest is slow/soft songs.

\textbf{Setup}: We used \textit{earbuds} for music streaming and IMU/audio sensing, and an Android \textit{phone} for connecting earbuds via Bluetooth and controlling music playing. For earbuds, we used Apple AirPods Pro, but also additionally used eSense~\cite{kawsar2018earables} to collect IMU data, i.e., one eSense unit on the left ear and one AirPod Pro unit on the right ear; note that AirPods Pro did not allow developers to access raw IMU data when we collected the data, but recently started to provide the access from iOS 14.0. The sampling rate of a microphone on AirPods Pro and IMU on eSense was set to 44.1 kHz and 70 Hz, respectively. For ground truth tagging, we recorded the data collection session with a covered camera, which is marked with a red rectangle in Figure~\ref{fig:data_collection}.

Note that we further conducted a small deployment study where participants used GrooveMeter without a camera (see Section~\ref{subsec:casestudy}).

%% file: sections/07.evaluation.tex
\section{Evaluation}

For evaluation, we implemented the prototype of GrooveMeter as an Android service on two phones, Galaxy S21 (Android 11.0) and Galaxy S8+ (Android 9.0). We use TensorFlow Lite~\cite{tflite} to run our pipelines with YAMNet and LSTM. We also implemented a prototype of a music player application that features a reaction summary and music recommendation, as shown in Figure~\ref{fig:gm_app}.

\input{sections/071.vocal_performance}

\input{sections/072.motion_performance}

\subsection{System Cost~\label{subsec:processingcost}}

\begin{table}[t]

  \begin{varwidth}[b]{0.4\linewidth}
    \footnotesize
    \centering
    \captionof{table}{System cost; time (ms), Power (mW).}
    \begin{tabular}{c|c|c|c|c|c}
    
    \hline
        &   & \multicolumn{2}{c|}{\textbf{GALAXY S21}}    & \multicolumn{2}{c}{\textbf{GALAXY S8+}}    \\ \cline{3-6}
                            & Operation                & Time & Power & Time & Power \\ \hhline{=|=|=|=|=|=}
    \multirow{4}{*}{\textbf{Vocal}} & Filtering      & 1.2      & 247.0     & 3.4       & 248.8      \\ \cline{2-6} 
                                    & Classification & 14.1     & 417.2    & 39.4      & 630.9    \\ \cline{2-6} 
                                    & Correction     & 1.4      & 261.9      & 10.9   & 275.8        \\ \cline{2-6} 
                                    & Smoothing      & 0.1      & 27.1     & 0.3        & 152.0     \\ \hhline{=|=|=|=|=|=}
    \multirow{3}{*}{\textbf{Motion}} & Filtering      & 0.5        & 27.9     & 1.2       & 28.4   \\ \cline{2-6} 
                                     & Classification & 12.4          & 92.62    & 34.6      & 82.3 \\ \cline{2-6} \hhline{=|=|=|=|=|=}
    \multirow{2}{*}{\textbf{Energy}}    & All (w/o filtering)      & \multicolumn{2}{c|}{7.3 mJ/s} & \multicolumn{2}{c}{70.4 mJ/s} \\ \cline{2-6}
    & All (w/ filtering) & \multicolumn{2}{c|}{3.7 mJ/s}  & \multicolumn{2}{c}{28.5 mJ/s} \\ \hline
    \end{tabular}
            
    \label{tab:system_cost}
    \end{varwidth}%
  \hspace{0.9in}
  \begin{minipage}[b]{0.4\linewidth}
    \centering
    \includegraphics[width=\columnwidth]{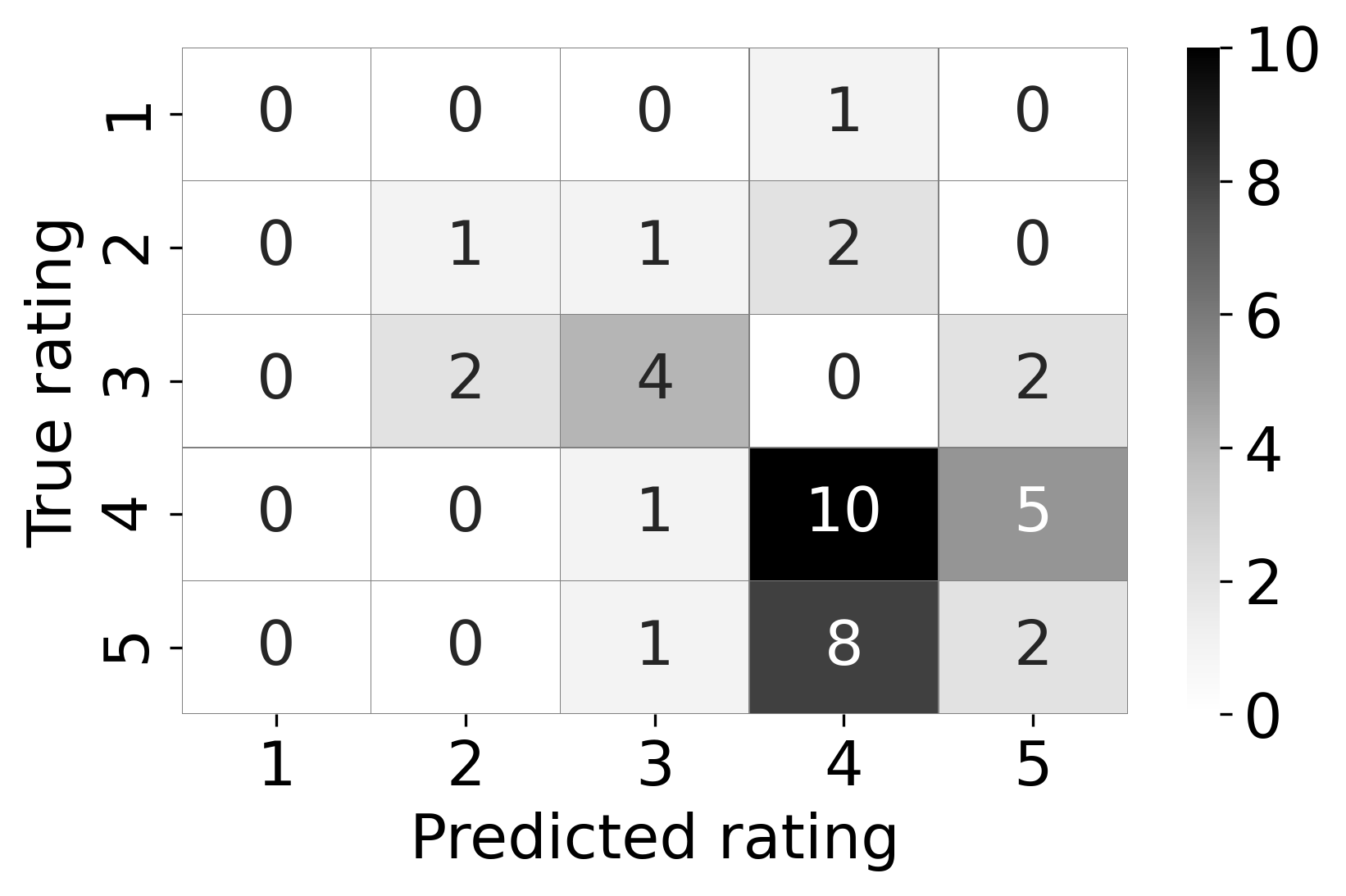}
    \vspace{-0.3in}
    \captionof{figure}{Confusion matrix of rating prediction.\label{fig:eval_rating}}
    
  \end{minipage}
  
\end{table}

We examine the system cost of GrooveMeter in terms of execution time and energy overhead.  Table~\ref{tab:system_cost} shows the results on two Android phones, Galaxy S21 and S8+. The end-to-end execution time of vocal and motion reaction detection components on Galaxy S21 is around \revise{17 ms and 13 ms}, respectively, which is suitable for supporting interactive applications. Even with the low-end phone, Galaxy S8+, the total time is still below than \revise{90 ms}. Note that the average latency is shorter than the total execution time because the classification and correction operations are not always performed due to the filtering operation.

We measure the energy cost using a Monsoon power monitor. The phones play music on Apple Airpod via Bluetooth and we measure net power increase by the operation execution. While the instantaneous power is high during the execution, the total energy cost with the filtering operation is marginal, i.e., \revise{3.7 mJ/s and 28.5 mJ/s} on Galaxy S21 and S8+, respectively. This is due to a) the energy piggybacking on the music player's use of CPU and communication and b) the filtering operation. Also, considering that GrooveMeter runs only when a user listens to music, the energy overhead would hardly impact the battery life.

\highlight{In addition, we examine the energy cost of the eSense earbuds. We use battery level information provided by Android since eSense does not provide an API to read the device's battery level. The information is coarse-grained with 10\% resolution. Our measurement for an hour shows that using eSense to listen to music with IMU sensing and audio recording consumes 20\% more battery than using eSense for music listening only. Note that the optimization of earbud energy consumption is our future work.}

\input{sections/074.application.tex}

%% file: sections/071.vocal_performance.tex
\subsection{Vocal Reaction Detection}

\begin{figure}[]
\begin{minipage}[t]{0.43\linewidth}
    \includegraphics[width=\linewidth]{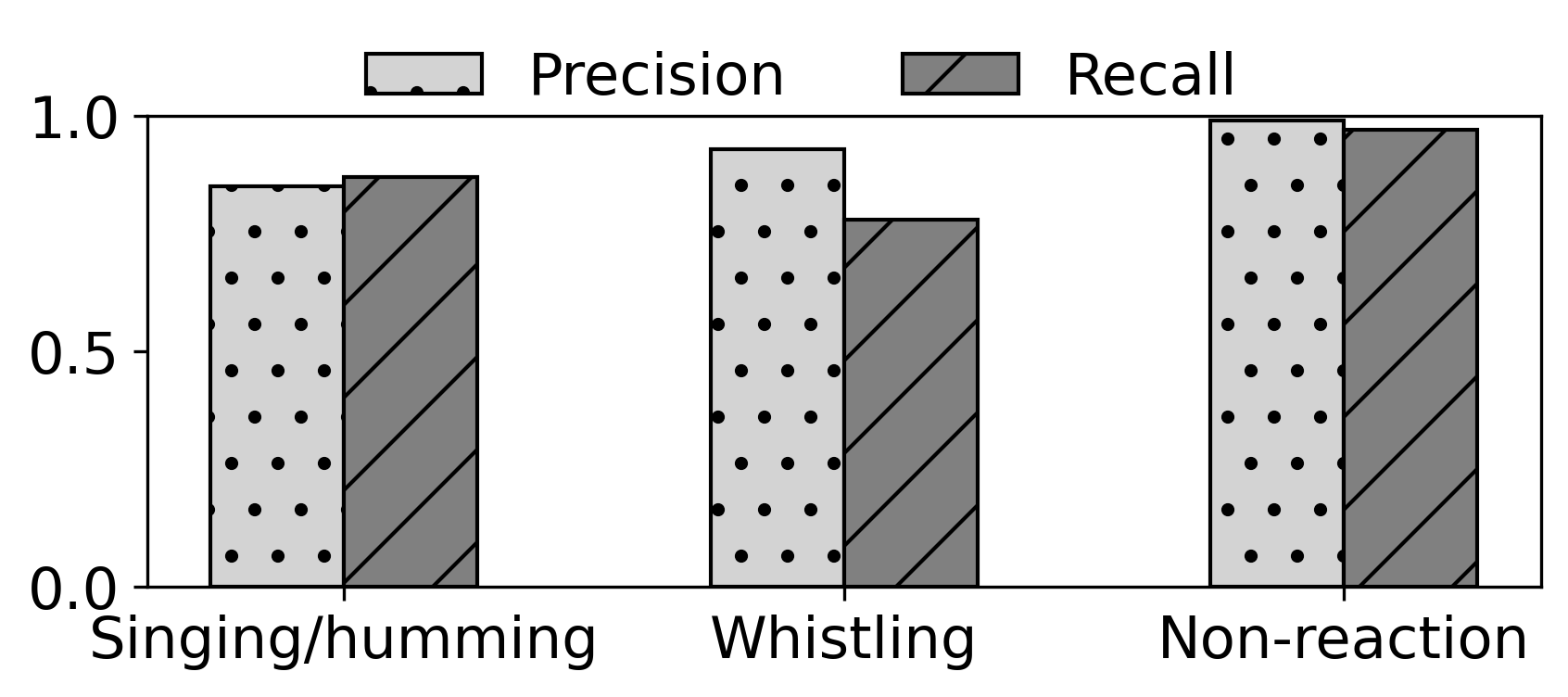}
    \vspace{-0.2in}  
      \caption{Overall performance}
      \label{fig:eval_vocal_overall}
    %   \vspace{-0.23in}  
    
\end{minipage}%
    \hspace{0.1in}
\begin{minipage}[t]{0.45\linewidth}
    \includegraphics[width=\linewidth]{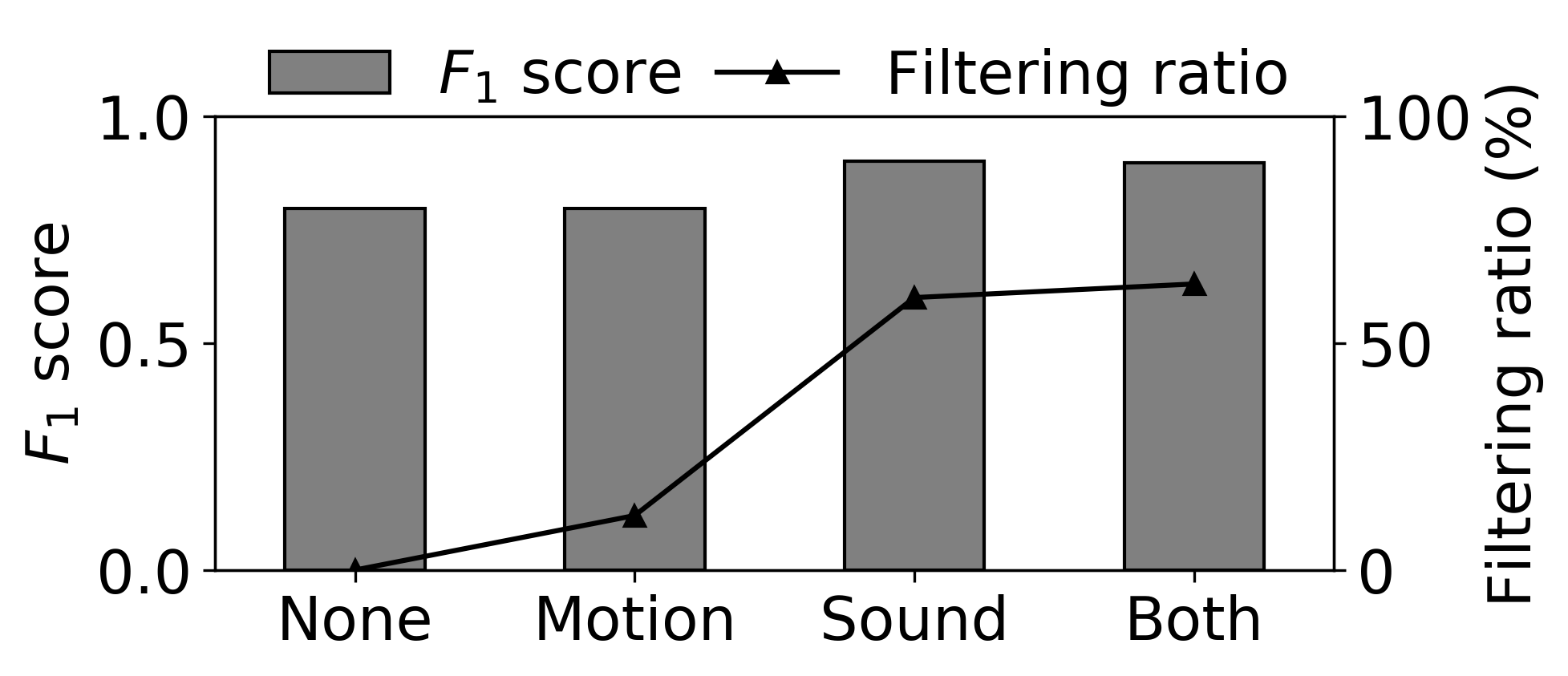}
    \vspace{-0.2in}  
    \caption{Effect of filtering}
    \label{fig:eval_f1_filtering_ratio}
    % \vspace{-0.23in}  
\end{minipage} 
\vspace{-0.1in}
\end{figure}

\subsubsection{Overall Performance}
%\hfill

We present the overall performance of the vocal reaction detection of GrooveMeter. For the validation, we used leave-one-subject-out (LOSO) cross validation (CV) with the MusicReactionSet dataset. Note that we used the original YAMNet model~\cite{yamnet} for the classification task, thus the same model is used for testing all subjects. LOSO CV is considered to obtain the threshold values in the vocal reaction pipelines (threshold ranges of motion and audio filters, pitch similarity threshold) to avoid the over-fitting problem.

Figure~\ref{fig:eval_vocal_overall} shows the averaged precision and recall of vocal reaction labels over \revise{30} validations. The experimental results show that GrooveMeter achieves the reasonable performance of the vocal reaction detection even for an unseen user and under a variety of real-life background audio noise types; the macro-averaged $F_1$ score is \revise{0.90}. More specifically, it detects \textit{singing/humming} reactions with \revise{0.85 and 0.87} of precision and recall, respectively, and \textit{whistling} reactions with \revise{0.93 and 0.78}. The recall of whistling is relatively low compared to other labels because some whistling segments with weak sound or mixed with background noise are incorrectly inferred by YAMNet. 
The results also show that our method correctly identifies non-reaction events. The precision and recall for the \textit{non-reaction} label are \revise{0.99 and 0.97}.

\subsubsection{Effect of filtering:} We examine the effect of the early-stage filtering. Figure~\ref{fig:eval_f1_filtering_ratio} shows the $F_1$ score and filtering ratio with different filtering strategies. For the study, we developed three different versions of our pipeline of the vocal reaction detection; \textit{none}, \textit{motion}, and \textit{sound}. We define the filtering ratio as the number of filtered segments divided by the total number of segments; i.e., a high filtering ratio means less processing cost. None means the voice reaction detection without any filtering operation, i.e., 0\% of the filtering ratio.
Motion and sound refer to the pipeline when only motion-based and sound-based operation is added, respectively. 

Interestingly, the filtering operation is not only effective for reducing computation cost, but also helpful in improving vocal reaction detection performance by effectively filtering out reaction-irrelevant audio signals. While \textit{none} achieves 0.8 of $F_1$ score without any filtering, applying both filtering operations, \textit{both}, increases the $F_1$ score by 0.1. We also observe the different effects of filtering modes. The filtering ratios of motion and sound-based operation are \revise{12\% and 60\%}, respectively. The simple motion-based filtering contributes to reducing a fair amount of non-reaction data to process without compromising the performance. The sound-based filtering reduces even more amount of data. At the same time, it effectively removes false positives from original YAMNet due to background noise, thereby increasing the $F_1$ score. When both operations are used together in GrooveMeter, the ratio increases up to \revise{63\%}, implying that a better performance is achieved even with the classification of \revise{37\%} of segments, compared to \textit{none}.

\begin{figure}[]
\begin{minipage}[t]{0.43\linewidth}
    \includegraphics[width=\linewidth]{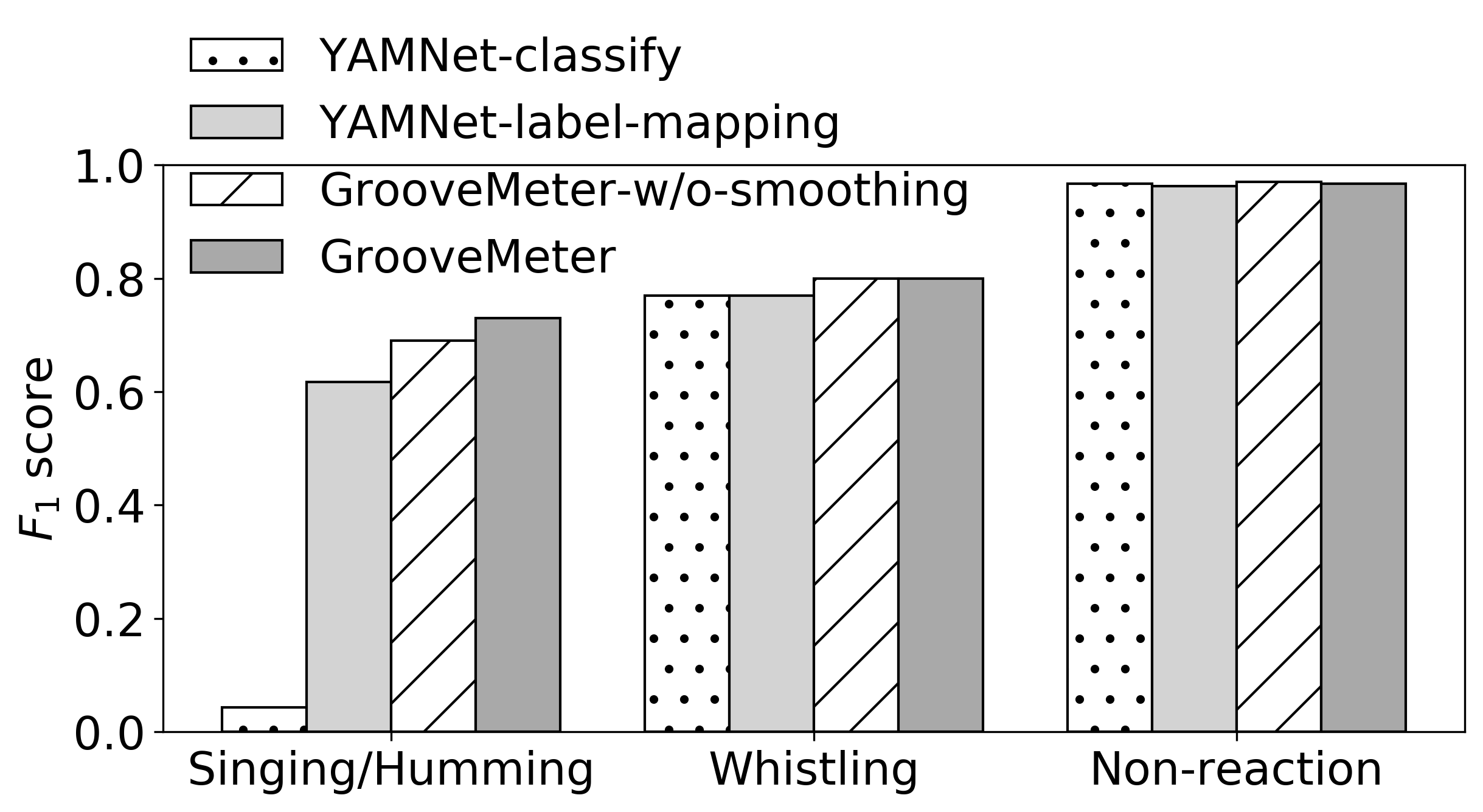}
    \vspace{-0.1in}
    \caption{Effect of correction and smoothing}
    \label{fig:eval_vocal_breakdown}
    
\end{minipage}%
    \hfill%
\begin{minipage}[t]{0.56\linewidth}
    \includegraphics[width=\linewidth]{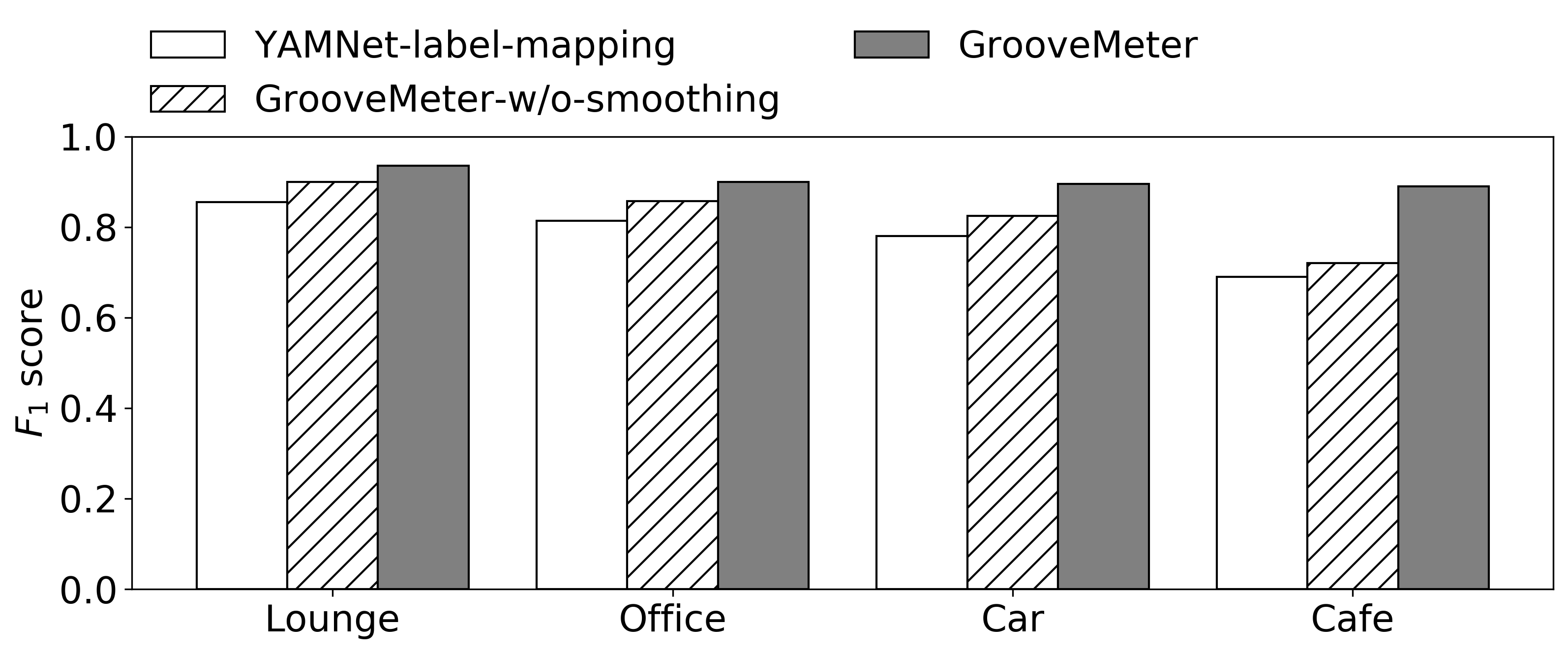}
    \vspace{-0.1in}
    \caption{Robustness against noise}
    \label{fig:eval_vocal_noise}
\end{minipage} 
\vspace{-0.1in}
\end{figure}

\subsubsection{Effect of correction and smoothing:} We further break down the performance of the rest of operations, correction and smoothing. To quantify the impact of each technique, we implemented three variants of our method, \textit{YAMNet-classify}, \textit{YAMNet-label-mapping} and \textit{GrooveMeter-w/o-smoothing} in which we apply each technique in turn, and compared their performance; here, we excluded the early-stage filtering operation to focus on the classification performance.
As a baseline, YAMNet-classify outputs reaction labels from the original YAMNet model without any further operation. YAMNet-label-mapping maps speech and music labels of the YAMNet model output to singing/humming. GrooveMeter-w/o-smoothing handles the ambiguous labels obtained from label mapping in Table~\ref{table:vocal_mapping} and rank constraint relaxation, and performs music information-leveraged correction, but without smoothing. 

Figure~\ref{fig:eval_vocal_breakdown} shows the $F_1$ score of the variants. Even trained with massive data, we observe that YAMNet-classify shows poor performance for vocal reaction data, especially for singing/humming (0.04 of $F_1$ score). As described in \S\ref{subsec:vocal_reaction}, this is mainly because the YAMNet model is extremely poor in distinguishing \textit{singing-along} data from \textit{speech} and \textit{music} data. The YAMNet model classifies only 1.8\% of singing/humming data as singing or humming labels. 
After applying labeling mapping, YAMNet-label-mapping shows meaningful improvement for the singing/humming data, but it still shows a considerable amount of false-positive errors, e.g., inferring background music and chat of nearby people as singing/humming. 
Our YAMNet output relaxation and music information-leveraged correction improve $F_1$ score compared to YAMNet-label-mapping. In particular, it improves the classification performance of vocal reactions (difference between YAMNet-label-mapping and GrooveMeter-w/o-smoothing). 
Specifically, the $F_1$ score of singing/humming increases from \revise{0.61 to 0.68. That of whistling increases from 0.77 to 0.8.} We also observe that HMM-based smoothing meaningfully improves the performance of vocal reaction detection (difference between GrooveMeter-w/o-smoothing and GrooveMeter). The $F_1$ score of singing/humming increases by \revise{0.06 from 0.68 and 0.74.} Note that the whistling is uncommon in our dataset, so the effect of smoothing is limited.

\subsubsection{Robustness against acoustic noise in real-life situations:}

We further investigate the robustness of our technique against noise in real-life situations. As presented in \S\ref{sec:collection}, we consider four places, \textit{lounge}, \textit{office}, \textit{car}, and \textit{cafe}, exhibiting different noise characteristics as shown in Table~\ref{table:data_situations}, and compare the detection performance in these places. We break down the performance by comparing GrooveMeter with two variants mentioned above: YAMNet-label-mapping and GrooveMeter-w/o-smoothing. Here, we include the early-stage filtering operation to examine overall performance.

Figure~\ref{fig:eval_vocal_noise} shows the macro-averaged $F_1$ scores for vocal reactions and non-reactions.
%Figure~\ref{fig:eval_vocal_noise} shows the average $F_1$ scores for singing/humming reactions in four places. 
The results show the robust detection performance of GrooveMeter regardless of different noise characteristics in four places. Compared to YAMNet-label-mapping (YAMNet-based classification and label mapping), GrooveMeter increases the $F_1$ score by 0.08, 0.09, 0.12, and 0.21 in lounge, office, car, and cafe, respectively, by adopting music information-leveraged correction (GrooveMeter-w/o-smoothing) and smoothing (GrooveMeter). The correction operation does contribute much in the cafe due to relatively large false positives from background noise (0.04 increase of $F_1$ score). However, interestingly, the smoothing operation shows meaningful improvement by leveraging the temporal association of reaction labels (further 0.17 increase of $F_1$ score), thereby achieving comparable performance to other places.

%% file: sections/072.motion_performance.tex
\subsection{Motion Reaction Detection~\label{subsec:motion_performance}}

We present the performance of our motion reaction detection. For the validation, we used LOSO CV with the MusicReactionSet dataset. We implement three baselines by referring to prior works as follows.

\begin{itemize}[noitemsep, topsep=4pt, leftmargin=*]
    \item RandomForest: It represents sensing pipelines to recognize repetitive and periodical physical activities, e.g., ~\cite{morris2014recofit, bedri2017earbit, khurana2018gymcam}. The pipelines are typically composed of feature extractors and machine learning classifiers. We use time and frequency-domain statistical features~\cite{figo2010preprocessing} and auto-correlation-derived features~\cite{bedri2017earbit, khurana2018gymcam} from IMU data. We tested the performance with popular classifiers such as support vector machine (SVM), logistic regression, decision tree, and random forest, and we chose the RandomForest model that outperforms the rest of classifiers.
    
    \item CNN: It represents a deep learning based method for human activity recognition, e.g.,~\cite{ronao2016human}, which uses convolutional neural network with 3 axes of accelerometer and 3 axes of gyroscope from IMU sensor stream. \highlight{We build a classification model consisting of 3 convolutional layers with ReLU activation function, max pooling layer, 2 dropout layers with a drop rate of 0.5, and a softmax layer.}

    \item ConvLSTM: It represents a deep learning based method that combines convolutional and LSTM recurrent layers for activity recognition with wearables, e,g.,~\cite{ordonez2016deep}. It uses convLSTM~\cite{shi2015convolutional} with 3 axes of accelerometer and 3 axes of gyroscope from IMU sensor stream. \highlight{We build a classification model consisting of a ConvLSTM layer, a dropout layer with a drop rate of 0.5, a ReLU layer, and a softmax layer.}

\end{itemize}

\subsubsection{Overall Performance}

\begin{figure}[]
    \centering
    \mbox{
        \subfloat[Comparison with baselines\label{fig:com_baseline}]
            {\includegraphics[width=0.35\columnwidth]{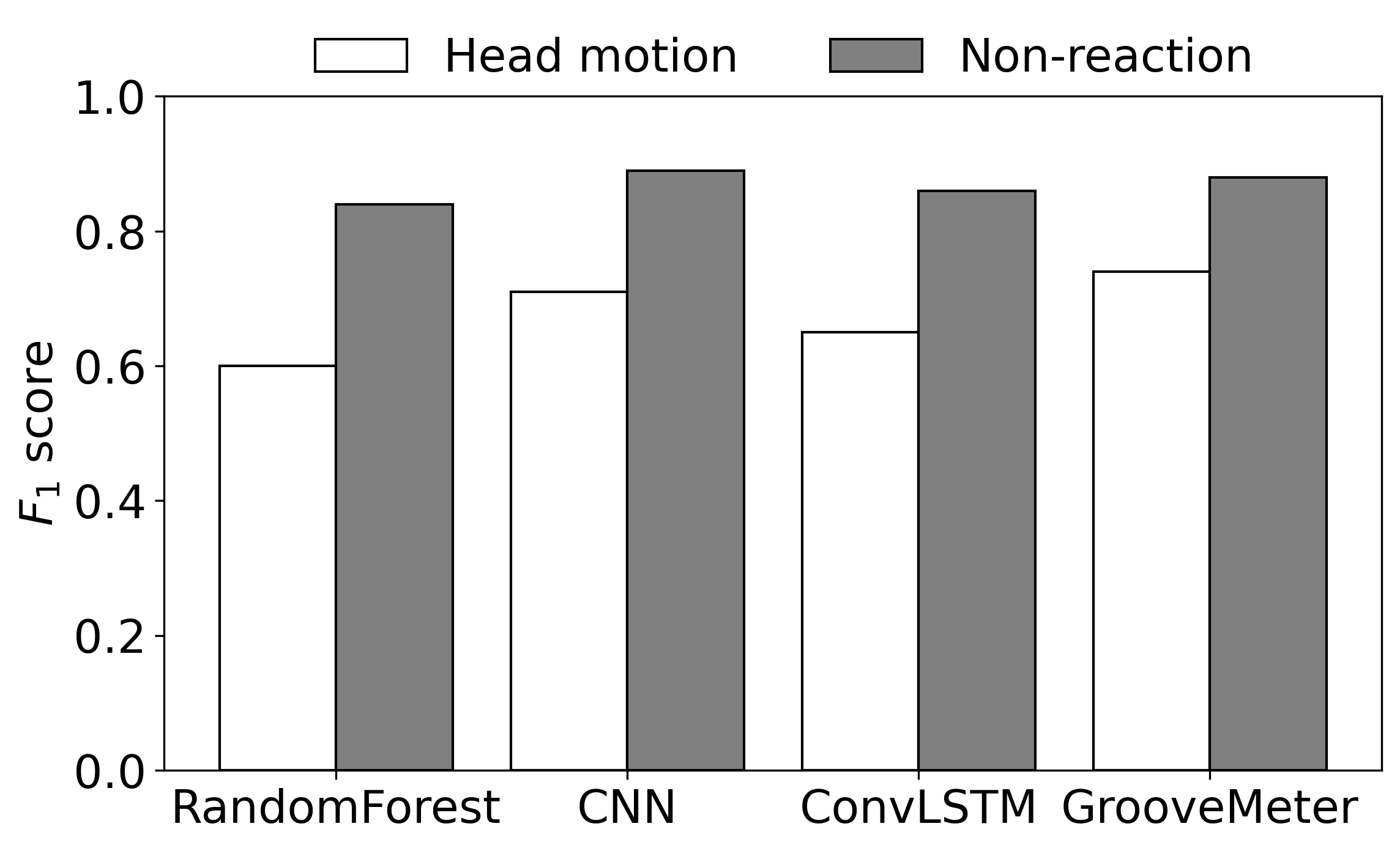}}
        \hspace{0.1in}

        \subfloat[Precision/Recall (GrooveMeter)\label{fig:motion_overall}]{
            \includegraphics[width=0.29\columnwidth]{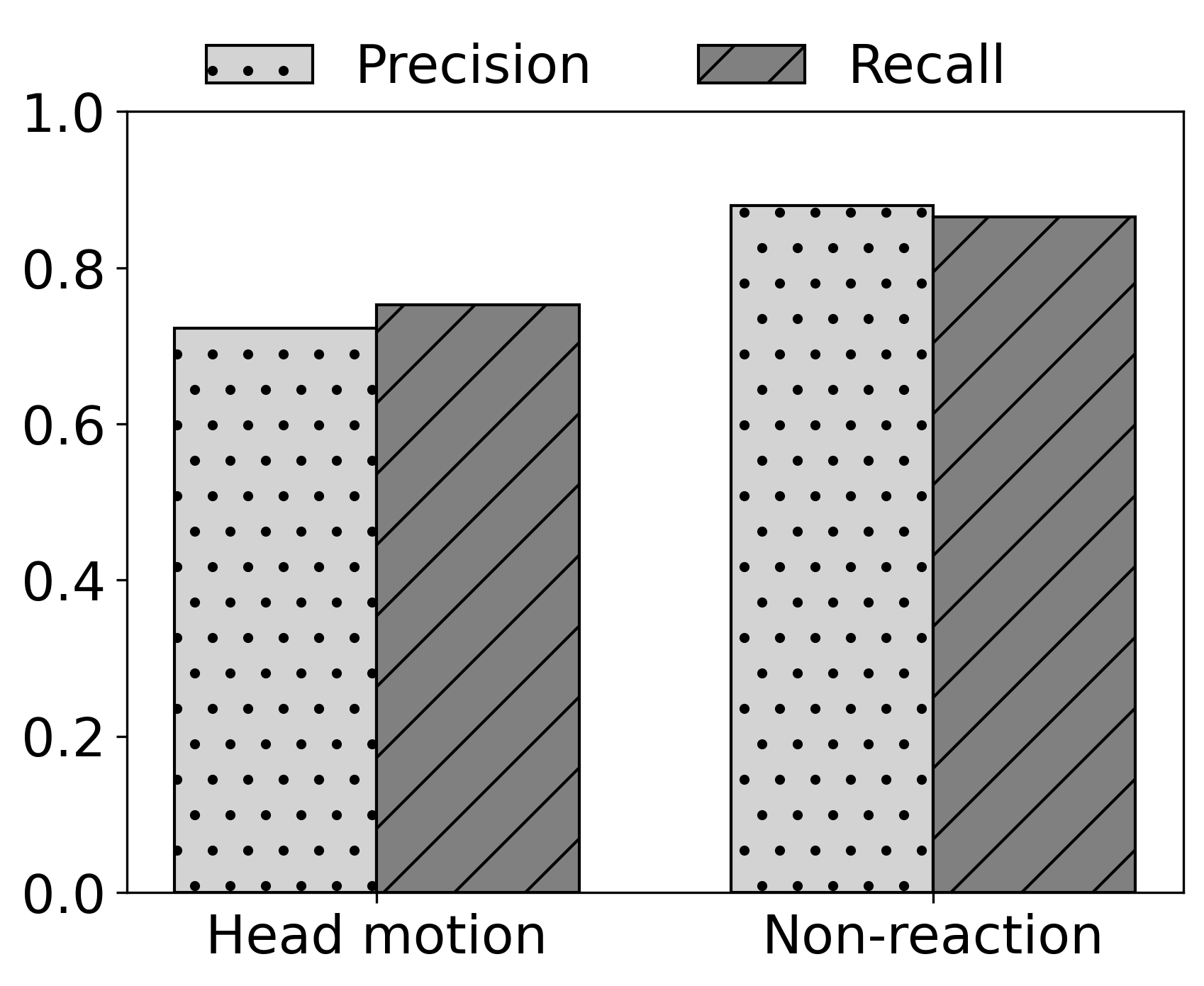}}
        \hspace{0.1in}
        
        \subfloat[Per-genre result (GrooveMeter)\label{fig:motion_genre}]
            {\includegraphics[width=0.29\columnwidth]{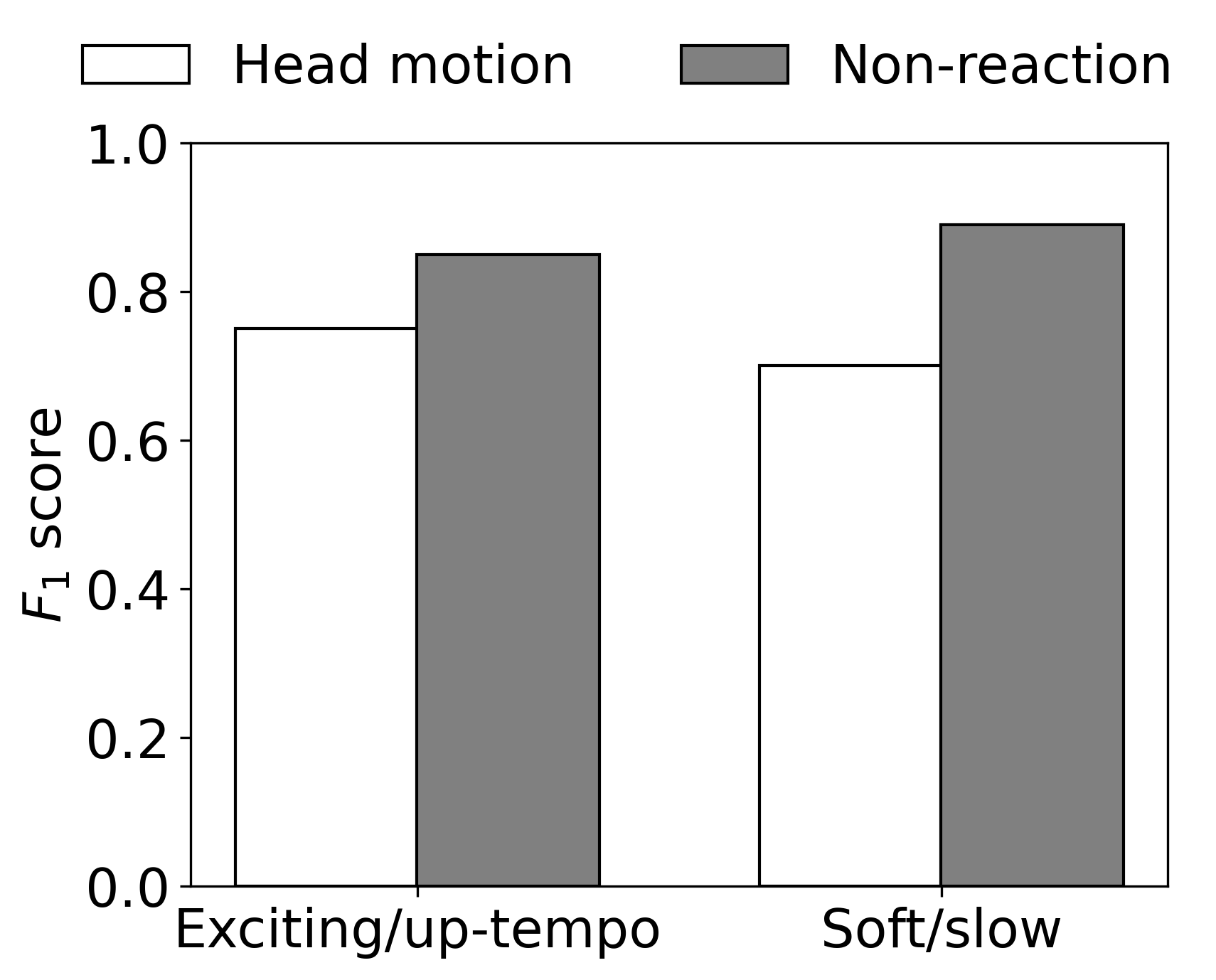}}
        }
    
    \caption{Motion reaction detection performance
    \label{fig:motion_reaction_overall}}
\vspace{-0.1in}
\end{figure}

Figure~\ref{fig:com_baseline} shows the performance comparison between GrooveMeter and the baselines. The results show that GrooveMeter detects the head motion more accurately than the baselines. While there is a marginal difference in the $F_1$ score of the non-reaction class, GrooveMeter increases the $F_1$ score of the head motion class by 0.09 on average, compared to the baselines. For the head motion, the $F_1$ score of GrooveMeter is 0.74, whereas that of RandomForest, CNN, and ConvLSTM is 0.60, 0.71, and 0.65, respectively. One may argue that the performance improvement of GrooveMeter from CNN (0.03 increase) is marginal. However, we found out that GrooveMeter is more robust to motion noisy environments. We present the in-depth analysis in \S\ref{subsub:motion_comparison}.

We then take a deeper look at the performance of GrooveMeter. Figure~\ref{fig:motion_overall} shows that, even for a unseen user, our method shows the reasonable performance of the motion reaction detection. Specifically, it detects \textit{head motion} with \revise{0.72} and \revise{0.75} of precision and recall, respectively. For \textit{non-reaction}, it achieves \revise{0.88} of precision and \revise{0.87} of recall. \cm{any more analysis?}

We look into the results depending on the genre (see Figure~\ref{fig:motion_genre}). \highlight{Note that we classify dance, fast-beat hip-hop/rap, and fast rock as up-tempo/exciting songs, and ballads, R\&B/soul, slow-beat rock, and folk/blues as slow/soft songs.} The results of exciting/up-tempo songs show higher precision and recall for the reaction than those of soft/slow songs, resulting in a relatively large $F_1$ score. \highlight{Specifically, the $F_1$ scores of \textit{head motion} and \textit{non-reaction} are 0.75 and 0.85, respectively, for exciting/up-tempo songs. Those for slow songs are 0.70 and 0.89, respectively.} While listening to up-tempo songs, people tend to nod vigorously. Thus, motion reaction shows more prominent signals and clear periodicity, which yields better performance. In contrast, with soft/slow songs, motion reactions tend to be rather weak, and their trajectory is also small. Thus, more reaction data can be confused with non-reaction motions compared to the up-tempo case.

\subsubsection{In-depth Comparison with Baselines~\label{subsub:motion_comparison}}

\begin{table}[t]
\footnotesize
\centering
\caption{Comparison with the baselines~\label{tab:comparison_baseline}} 
 
\begin{tabular}{c c c c c c c c c c c c c}
\hline
    &  \multicolumn{2}{c|}{\textbf{Controlled}} &\multicolumn{10}{c}{\textbf{MusicReactionSet}} \\ \cline{4-13} 
              &     \multicolumn{2}{c|}{\textbf{}} &  \multicolumn{2}{c}{{Lounge}}& \multicolumn{2}{c}{{Office}}& \multicolumn{2}{c}{{Car}}& \multicolumn{2}{c|}{{Cafe}}& \multicolumn{2}{c}{{Avg.}} 
                                      \\ \hhline{ = = = = = = = = = = = = =}
{\textbf{Head motion}}      &      &     &       &     &     &     &      &   \\ \cline{1-13} 
{{RandomForest}}      & \multicolumn{2}{c}{{0.96}}    & \multicolumn{2}{c}{{0.61}}   & \multicolumn{2}{c}{{0.54}} & \multicolumn{2}{c}{{0.67}} &\multicolumn{2}{c}{{0.59}}  &\multicolumn{2}{c}{{0.60}} \\ 
{{CNN}}      & \multicolumn{2}{c}{{0.93}}    & \multicolumn{2}{c}{{0.74}}   & \multicolumn{2}{c}{{0.68}} & \multicolumn{2}{c}{{0.75}} &\multicolumn{2}{c}{{0.66}} &\multicolumn{2}{c}{{0.71}}\\
{{ConvLSTM}}      & \multicolumn{2}{c}{{0.97}}    & \multicolumn{2}{c}{{0.66}}   & \multicolumn{2}{c}{{0.57}} & \multicolumn{2}{c}{{0.73}} &\multicolumn{2}{c}{{0.64}} &\multicolumn{2}{c}{{0.65}}\\
{{GrooveMeter}} &\multicolumn{2}{c}{{0.94}}  & \multicolumn{2}{c}{{0.75}}    & \multicolumn{2}{c}{{0.72}}   & \multicolumn{2}{c}{{0.75}} & \multicolumn{2}{c}{{0.72}}   &\multicolumn{2}{c}{{0.74}}  \\ \cline{1-13} 

{\textbf{Non-reaction}}        &      &     &       &     &     &     &      &  \\ \cline{1-13} 
{{RandomForest}}     & \multicolumn{2}{c}{{0.92}}    & \multicolumn{2}{c}{{0.81}}   & \multicolumn{2}{c}{{0.90}} & \multicolumn{2}{c}{{0.80}} &\multicolumn{2}{c}{{0.86}}  &\multicolumn{2}{c}{{0.84}}  \\ 
{{CNN}}      & \multicolumn{2}{c}{{0.85}}    & \multicolumn{2}{c}{{0.84}}   & \multicolumn{2}{c}{{0.92}} & \multicolumn{2}{c}{{0.86}} &\multicolumn{2}{c}{{0.89}} & \multicolumn{2}{c}{{0.89}}\\
{{ConvLSTM}}      & \multicolumn{2}{c}{{0.92}}    & \multicolumn{2}{c}{{0.81}}   & \multicolumn{2}{c}{{0.89}} & \multicolumn{2}{c}{{0.86}} &\multicolumn{2}{c}{{0.89}} &\multicolumn{2}{c}{{0.86}}\\
{{GrooveMeter}}     & \multicolumn{2}{c}{{0.86}}    & \multicolumn{2}{c}{{0.84}}   & \multicolumn{2}{c}{{0.91}} & \multicolumn{2}{c}{{0.85}} &\multicolumn{2}{c}{{0.90}} &\multicolumn{2}{c}{{0.88}}   \\ \cline{1-13} 

\end{tabular}

\end{table}

To better understand the performance difference with the baselines, we additionally collected the \textit{controlled} dataset in a controlled lab setting environment. We recruited 10 participants (M: 6, F: 4) from a university campus, $P_B1$ to $P_B10$; their ages are 20-26 (mean: 23.1). They all voluntarily participated and were compensated with a gift card worth USD 9. We invited each participant to the lab and asked to follow an instructed scenario including 3 sessions: 2 sessions to collect music listening reactions and 1 for others. In the first 2 sessions, we provided the top 100 music chart in a music streaming service and let them freely select a song to listen to in every session. Then, they were asked to make a given reaction naturally, but continuously for 60 to 90 seconds. The last session's task was freely moving around the lab and making music-irrelevant motions but while listening to music, which represent \textit{non-reaction}.

Table~\ref{tab:comparison_baseline} shows the $F_1$ scores of \textit{head motion} and \textit{non-reaction}, respectively, in the controlled and MusicReactionSet datasets. We first look into the detection performance of the head motion. Interestingly, while all the methods including GrooveMeter show the similar performance in the controlled dataset, the performance difference is noticeable in the MusicReactionSet data (which was collected in the wild setting). For example, the $F_1$ scores for the head motion are over 0.93 in the controlled dataset. In MusicReactionSet, the $F_1$ score generally decreases due to various motion noises and diverse motion reaction patterns, but GrooveMeter shows the smaller gap compared to the baselines. 

We further examine the head motion detection in different situations. One interesting observation is that, while GrooveMeter shows the similar performance across the situations, the performance of all the baselines decreases much in the office and cafe situations where more motion noises are observed, as described in Table~\ref{table:data_situations}. For example, CNN shows the similar average $F_1$ score of the head motion (0.71) to GrooveMeter (0.74), its decrease in $F_1$ scores in the office and cafe situations ranges from 0.06 to 0.09 compared to lounge and car situations. On the contrary, the decrease of GrooveMeter is just 0.03, which shows that GrooveMeter is much more robust to daily motion noises.

We look into the detection performance of the non-reaction class. In the controlled data, CNN and GrooveMeter shows much lower $F_1$ scores than RandomForest and ConvLSTM, but they show higher performance in the MusicReactionSet. In MusicReactionSet, for non-reaction, the $F_1$ scores of CNN and GrooveMeter are 0.89 and 0.88, whereas those of RandomForest and ConvLSTM are 0.84 and 0.86. We conjecture that this is because the head motion and non-reaction segments have the clearly distinguishable pattern in the controlled data. However, in MotionReactionSet, there are more confusing cases due to various patterns of natural head motions and daily motion noises, thereby increasing false positive errors of RandomForest and ConvLSTM.

\begin{figure}[t!]
    \centering
    \includegraphics[width=0.7\columnwidth]{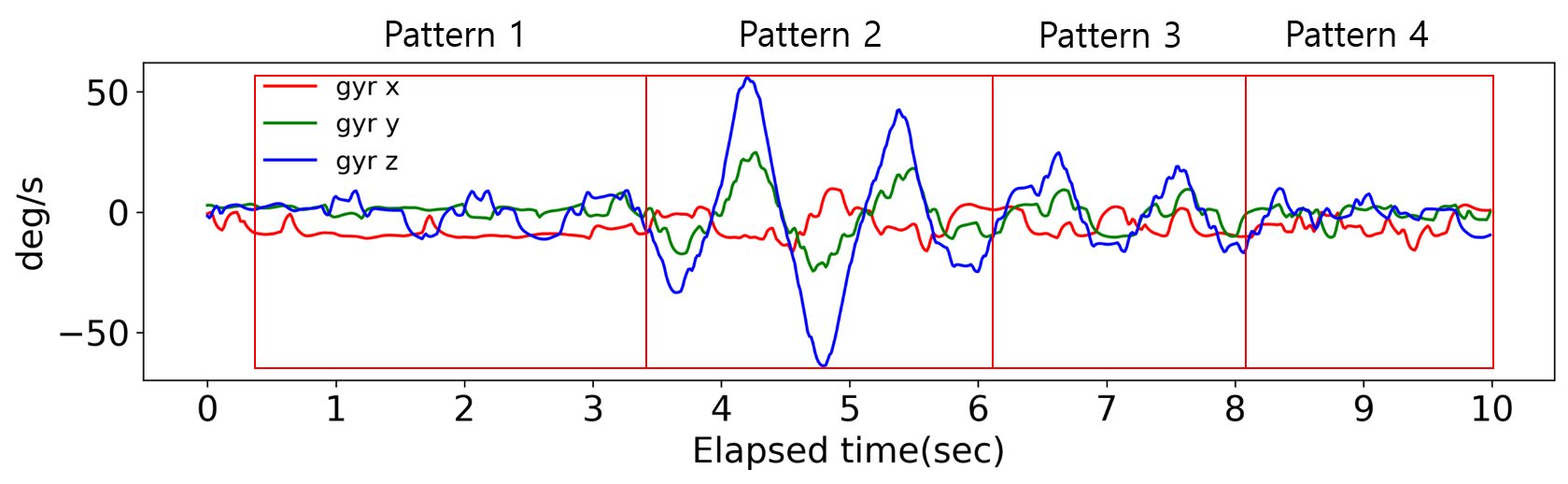}
    
    \caption{Gyroscope data with head motion over 10 seconds. \highlight{Pattern 1: Swaying head very slowly. Pattern 2-4: Nodding head vigorously and then gradually weakly}\label{fig:various_motion}}
\vspace{-0.15in}    
\end{figure}

We showcase a representative example that explains why GrooveMeter is more accurate and robust to the head motion detection than the baselines. Figure~\ref{fig:various_motion} depicts the 10 second-long gyroscope signal where the head motion is continuously performed. As shown in the figure, even in one session of the head motion, the movement pattern (i.e., corresponding signal pattern) is not regular or uniformly repetitive; the direction and strength of head motion naturally changes at short intervals. The existing motion sensing pipeline (e.g., RandomForest) could fail to detect such cases because they usually assume that the statistical features are similar over the window size, e.g., 7 seconds. On the other hand, GrooveMeter leverages the temporal sequence of motion units (100 ms) using the LSTM classifier for the head motion detection, thereby being able to handle various pattern of head motions.

\subsubsection{Effect of Activities}

\begin{figure}[]
\begin{minipage}[t]{0.47\linewidth}
    \includegraphics[width=\linewidth]{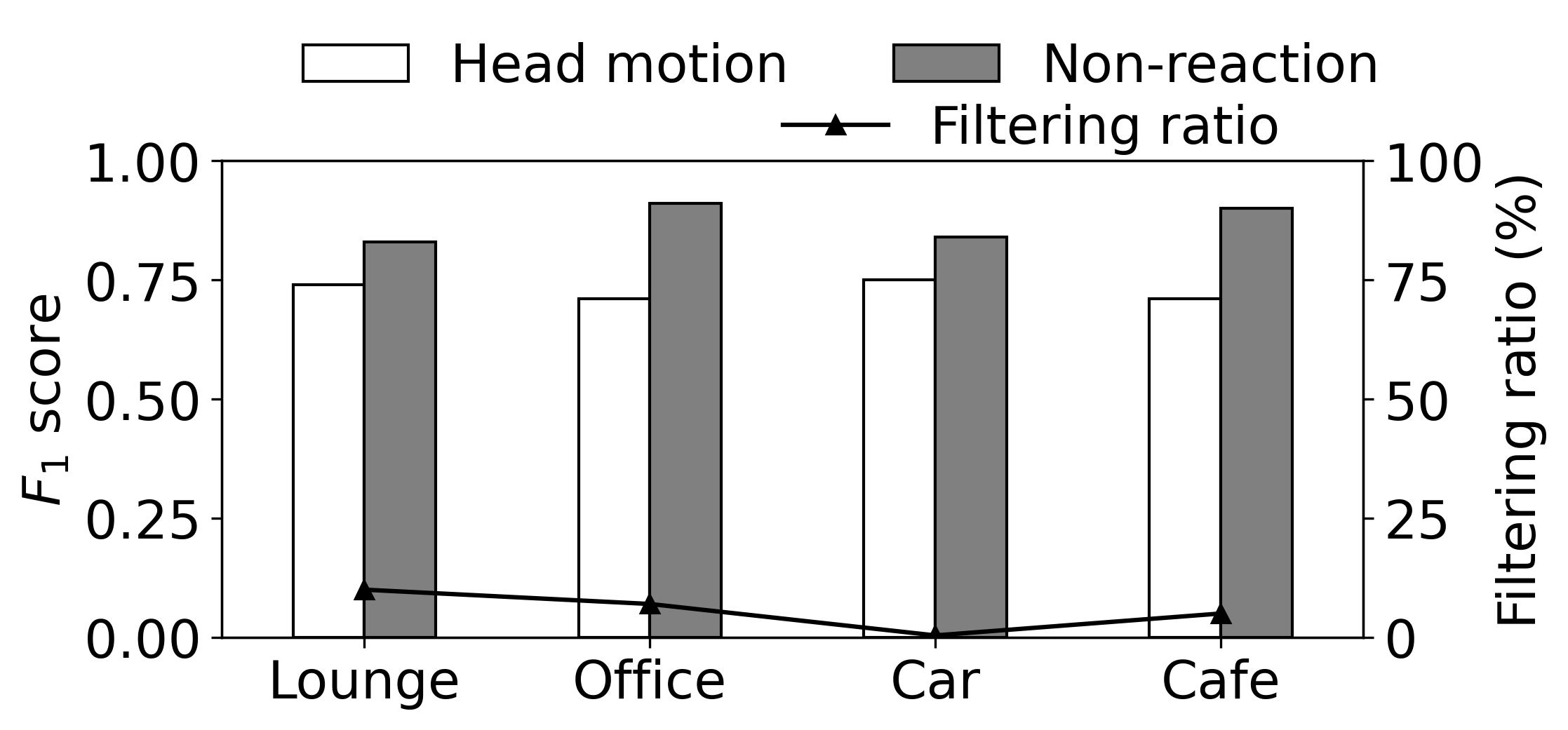}
    \vspace{-0.1in}
    \caption{Effect of activities in different places~\label{fig:eval_motion_noise}}
    
\end{minipage}%
    \hspace{0.25in}
\begin{minipage}[t]{0.38\linewidth}
    \includegraphics[width=\linewidth]{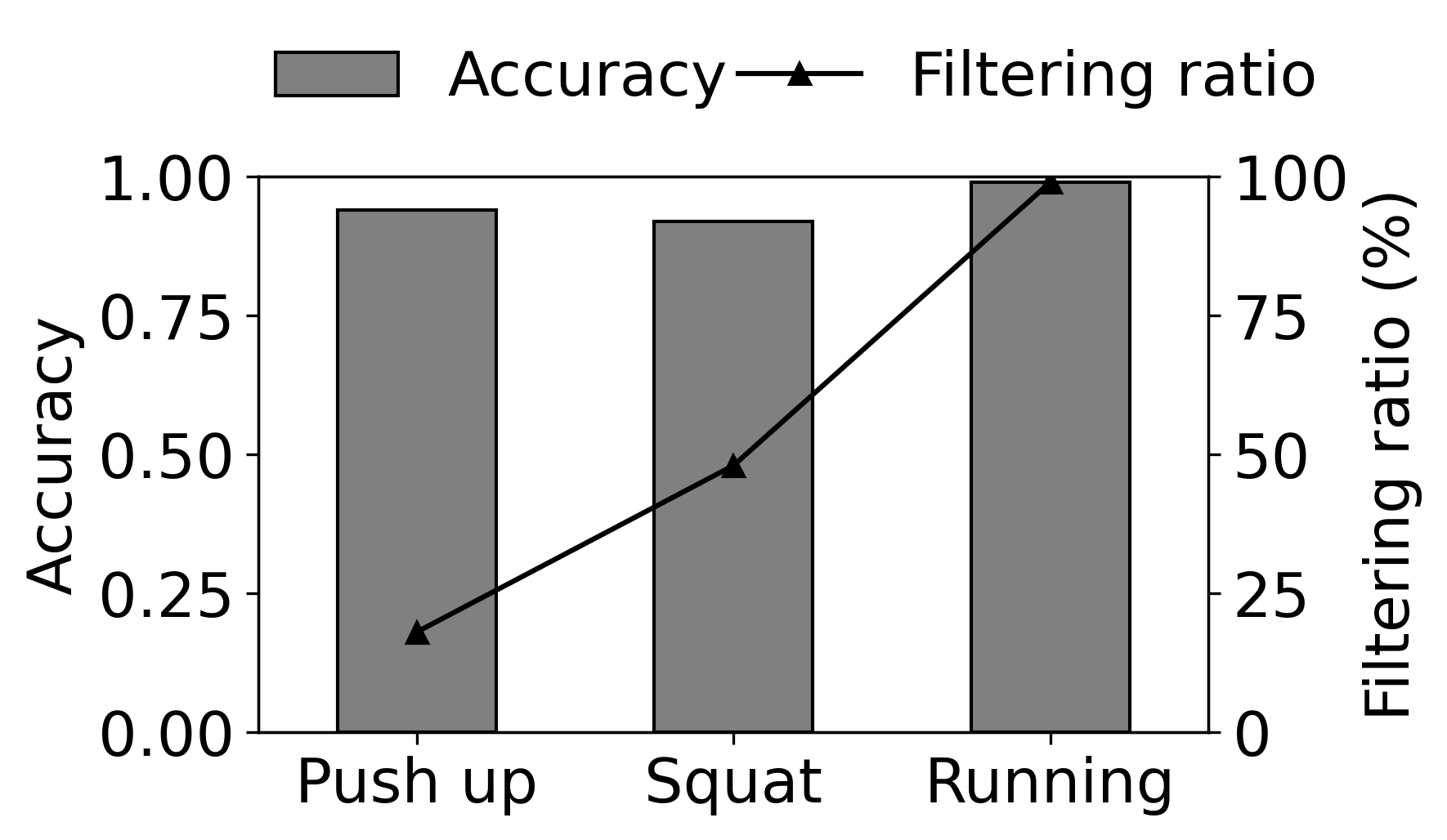}
    \vspace{-0.1in}
    \caption{Robustness against other activities~\label{fig:eval_motion_excise_noise}}
\end{minipage} 
\vspace{-0.1in}
\end{figure}

We examine the performance of motion reaction detection depending on the activities. As mentioned in \S\ref{sec:collection}, the participants did different activities in different places, i.e., resting in a lounge, working at an office, riding in a car, and relaxing at a cafe, which cause various reaction-irrelevant movements. 

Figure~\ref{fig:eval_motion_noise} shows the macro-averaged $F_1$ scores and filtering ratios of four cases. Regardless of activities, GrooveMeter achieves similar performance. The $F_1$ scores of head motion in the office and cafe cases are slightly smaller than the others. On the contrary, the non-reaction's $F_1$ scores are opposite. The filtering ratios are different from each other. That of lounge is 10\%, larger than the others. The ratio of car is only 0.4\% since most of non-reaction data are within the filter range due to the movement by the driving car.

\subsubsection{Robustness against Other Activities}

We further investigate the robustness in real-life situations where people often listen to music while doing other activities such as exercises. For the purpose, we additionally collected IMU data from 5 participants in the same way described in \S\ref{sec:collection}. While listening to three songs they liked, they did three activities, i.e., \textit{running}, \textit{doing push-ups}, and \textit{doing squat}. 
They did not make any reactions while doing exercises.
To examine the detection performance, we train the classification model only with the dataset with 30 participants in \S\ref{sec:collection}, and use the other activity data for test. We use accuracy since this includes non-reaction data only. Figure~\ref{fig:eval_motion_excise_noise} shows that GrooveMeter is robust against the motions involved in the three activities. The accuracy are 0.94, 0.92, and 0.99 for push up, squat, and running, respectively. Although the activities exhibit periodic movements, GrooveMeter accurately distinguishes non-reactions. Also, except the push up case, the filtering ratios are high, 48\% and 99\%, which significantly reduces processing overhead. Still, the push up case shows 18\% of filtering ratio which can reduce a fair amount of processing cost.

%% file: sections/074.application.tex
\subsection{Application Case Study~\label{subsec:casestudy}}

We present the potential of GrooveMeter by conducting application case studies. Note that we neither argue the general performance characteristics nor the novelty of the techniques. We examine the feasibility and potential of music listening reaction information under interesting application scenarios.

\subsubsection{Experience Capturing}

For the study, we further recruited ten participants ($P_C1$ to $P_C10$) and asked them to listen to eight songs; four songs were randomly selected from the top 50 chart (which they never listened to) and the other four songs were chosen from the participants' playlist.

\textbf{Automatic music rating:} We study the feasibility of \textit{automatic music rating} based on vocal and motion reactions. For the study, we asked the participants to provide a music rating for eight songs they listened to, on a scale of 5; 5 means \textit{"I like this song very much."}. To build a rating prediction model, we extract statistical features from the detection output of vocal and motion reactions, i.e., normalized duration and number of each reaction label, and use a decision tree as a classifier. We examine the prediction performance of a rating with LOSO CV. For training and testing, we only used the data from unknown songs from the top 50 chart for two reasons. First, the ratings of known songs are skewed to 4 and 5 because they chose \textit{known} songs from their own playlist which may be composed of their favorite songs. Second, the rating prediction is useful for unknown songs because the rating for known songs is more likely to be already available.
Figure~\ref{fig:eval_rating} shows the confusion matrix of our rating prediction. Higher values around the diagonal indicate that predicted ratings based on the reactions are meaningfully close to the actual ratings (MAE: 0.22). 

\textbf{Familiarity detection:} We investigate if the familiarity of a song can be detected using music listening reactions, i.e., to detect if a user has already listened to a song before or not. This functionality would help music streaming services accelerate to build a new subscriber's music preference without explicitly asking which songs have been enjoyed before. We build a decision-tree model trained by using statistical features of vocal and motion reaction events. Similarly to automatic music rating, we validate its performance in a LOSO manner, but using the full dataset. The $F_1$ score for the detection of known and unknown songs is 0.78 (precision: 0.85, recall: 0.72) and 0.81 (precision: 0.76, recall: 0.88), respectively.

\subsubsection{Small-scale Deployment Experience}

We conduct additional experiments to observe GrooveMeter’s detection performance and recommendation feasibility under uncontrolled, natural music listening situations. For the study, we developed an application with two features, one to investigate the performance of reaction detection and the other to study reaction-based music recommendation as shown in Figure~\ref{fig:reaction_detection_screenshot} and ~\ref{fig:recommendation_screenshot}, respectively. We recruited three participants ($P_D1$ to $P_D3$) and installed GrooveMeter and these applications on their Android phones. They freely used GrooveMeter for music listening for one day. They listened to 21.6 songs on average. Then, we conducted semi-structured one-on-one interviews.

\textbf{Reaction-based music recommendation:} 
After playing a song, the application computes the reaction patterns and recommends a new song which shows the most similar reaction pattern out of the pool; we used 240 music listening sessions in the MusicReactionSet as a pool. We map labels to the reaction index (non-reaction: 0, singing: 1, whistling: 2, motion: 3) 
and define the reaction pattern as a sequence of 1-second-long indices and compute the similarity using DTW. 
Overall, the participants liked the recommended songs even though they never listened to those songs. Interestingly, they mentioned that, when they listened to the recommended song, they made a similar reaction pattern to the previous song. $P_D1$ said, \textit{"It recommended the style of songs I like. ... I think I also made similar reactions."}. $P_D3$ stated, \textit{"I liked the recommendation when I actually listened to."}. In common, they reported that reaction-based music recommendation looks useful. 

\textbf{Reaction detection in the wild:} To investigate the reaction detection in the wild, the application shows the time-series reaction output as shown in Figure~\ref{fig:reaction_detection_screenshot}, after a song is played. In the interview, we asked about their impression and perception of the reaction output. Note that this is a subjective, coarse-grained evaluation due to the lack of a camera for ground truth data, but is designed to evaluate reactions that users naturally made in totally uncontrolled settings. All of them stated that GrooveMeter showed a reasonable output of reactions according to their remembrance. $P_D2$ mentioned, \textit{"I thought it was fairly accurate."}. We asked them to rate the perceived detection accuracy on a scale of 5; 5 means "The summary looks very accurate", and the average score was 4. 

\textbf{Suggestions:} The participants suggested extended features of GrooveMeter based on their experiences. $P_D2$ and $P_D3$ mentioned that GrooveMeter could help music composers by indicating which part of a song listeners liked and reacted. $P_D2$ added, \textit{"It will be useful even in the concert if it shows when audiences went wild"} 

\highlight{While the current study shows the potential of real deployment use case, it is limited in terms of the number of participants and the time of use. Comprehensive user study with a longer term is our future work.}

%% file: sections/08.discussion.tex
\section{Discussion}

\textbf{Other types of reactions to music:} According to existing music psychology literature~\cite{hallam2016oxford}, reactions that people usually experience while listening to music are diverse. In addition to physical responses, they include physiological responses (e.g., heart rate, skin conductance, biochemical responses) and emotional responses (e.g., subjective feelings, emotional expressions). In this work, we focus on physical responses, and target vocal and motion reactions commonly observed in our in-the-wild dataset. GrooveMeter can be extended to detect physiological and emotional reactions by employing additional sensing modalities to earbuds, e.g., photoplethysmography (PPG)~\cite{ferlini2022ear}, skin conductivity (EDA)~\cite{pham2021detection}, and relevant recognition models. Moreover, it would be possible to distinguish additional types of physical responses such as foot tapping and finger snapping. We leave it as our future work.

\textbf{Threshold-based operations:} GrooveMeter includes several components that work based on the threshold values, e.g., threshold range of filtering and similarity threshold for correction. We chose such simple approaches instead of complex machine learning (ML)-based operations due to the following reasons. First, according to our preliminary analysis, threshold-based operations showed comparable performance to ML-based operations even without cost for model training and execution. Second, the threshold values were set in a conservative manner, e.g., threshold range of filtering operations was set to include reaction segments as many as possible, thus errors can cause the increase of system cost, but hardly affect the detection accuracy. However, we admit that threshold-based static operations would not be robust to diverse real-life situations. As future work, we will further investigate two approaches and explore online adaptation to make them robust even in unseen environments, i.e., dynamically adapting threshold values or ML models by reflecting runtime data. 

\textbf{Dataset:} Our dataset contains 240 music listening sessions from 30 participants in four real-life situations where people often enjoy listening to music. It allows a systematic understanding of signal characteristics of music listening reactions, but still has limitations in terms of surrounding environments, participant groups. Real-life music listening situations may be quite diverse, and the factors that can affect the detection performance might be different, e.g., the listener's reaction pattern, mobility pattern, noise conditions. It would be worth collecting a larger set of data covering diverse conditions and exploring to make the model more robust. 

\textbf{Context-dependent analysis:}
When listening to music while doing other primary activities (e.g., studying, reading, exercise), people may have less reaction than when listening to music is their primary activity. In addition, even if they listen to the same song, reactions to the song may change over time or vary depending on the circumstances. Providing additional information about users' context, e.g., activities, time and location, would be helpful. Moreover, context-dependent analysis would be one of the advanced functionalities of GrooveMeter in the future. % The implication of the reaction variation for diverse music engagement-aware applications needs further study. 

\textbf{Beyond music listening reaction:} Earbuds are primarily used for music listening, but they are also widely used in many daily-life situations, e.g., in online meetings, in online lectures, and even in face-to-face conversations for live translation. However, earbuds are still limited to audio streaming and lack understanding reactions that a user makes to the counterpart when earbuds are used for other purposes. We envision that GrooveMeter can be further extended to detect reactions in different situations and enrich context-awareness of these applications. For example, GrooveMeter can detect nonverbal and behavioral cues that students make in online lectures and help teachers instantly keep track of students' learning status. Similarly, a translation app can instantly adapt the translation result based on the reaction of a speaker or an opponent.

%% file: sections/09.conclusion.tex
\section{Conclusion}

We present GrooveMeter, a novel system to automatically detect vocal and motion reactions of music listeners via earable sensing. We devise the sophisticated processing pipelines to make reaction detection accurate, robust, and efficient. We present extensive experiments to evaluate GrooveMeter with a dataset containing 926-minute-long IMU and audio data with 30 participants in daily music-listening situations. We also demonstrate its usefulness through a case study.